\documentclass[structabstract]{aa}  
\usepackage{amssymb}
\usepackage{graphicx}
\usepackage{epsfig}
\usepackage{subfigure}
\usepackage[authoryear]{natbib}
\usepackage[figuresright]{rotating}

\usepackage{txfonts}
\newcommand{\msun}{$M_{\sun}$}

\newcommand{\teff}{$T_{\mathrm{eff}}$}
\newcommand{\logg}{$\log g$}
\newcommand{\lL}{$\log \frac{L}{L_{\odot}}$}
\newcommand{\mdot}{$\dot{M}$}
\newcommand{\myr}{$M_{\sun}$ yr$^{-1}$}
\newcommand{\vsini}{$V$~sin$i$}
\newcommand{\vinf}{$v_{\infty}$}

\newcommand{\vmac}{$v_{\mathrm mac}$}

\newcommand{\kms}{km s$^{-1}$}

\newcommand{\heii}{\ion{He}{ii} 4542}
\newcommand{\hei}{\ion{He}{i} 4471}

\begin{document}
   \title{A quantitative study of O stars in NGC2244 and the Mon OB2 association}

   \subtitle{}

   \author{F. Martins\inst{1}
          \and
          L. Mahy\inst{2}
          \and
          D.J. Hillier\inst{3}
          \and
          G. Rauw\inst{2}
          }

   \offprints{F. Martins}

   \institute{LUPM--UMR 5299, CNRS \& Universit\'e Montpellier II, Place Eug\`ene Bataillon, F-34095, Montpellier Cedex 05, France\\
              \email{fabrice.martins AT univ-montp2.fr}
         \and 
             Institut d'Astrophysique et de G\'eophysique, Universit\'e de Li\`ege, B\^at B5C, All\'ee du 6 Ao\^ut 17, B-4000, Li\'ege, Belgium\\
         \and
             Department of Physics and Astronomy, University of Pittsburgh, 3941 O'Hara street, Pittsburgh, PA 15260, USA \\
             }

   \date{Received ...; accepted ...}

 
  \abstract
   {}
   {Our goal is to determine the stellar and wind properties of seven O stars in the cluster NGC2244 and three O stars in the OB association MonOB2. These properties give us insight into the mass loss rates of O stars, allow us to
check the validity of rotational mixing in massive stars, and to better understand the effects of the ionizing flux and wind mechanical energy release on the surrounding interstellar medium and its influence on triggered star formation. }
   {We collect optical and UV spectra of the target stars which are analyzed by means of atmosphere models computed with the code CMFGEN. The spectra of binary stars are disentangled and the components are studied separately. }
   {All stars have an evolutionary age less than 5 million years, with the most massive stars being among the youngest. Nitrogen surface abundances show no clear relation with projected rotational velocities. Binaries and single stars show the same range of enrichment. This is attributed to the youth and/or wide separation of the binary systems in which the components have not (yet) experienced strong interaction. A clear trend of larger enrichment in higher luminosity objects is observed, consistent with what evolutionary models with rotation predict for a population of O stars at a given age. 
We confirm the weakness of winds in late O dwarfs. In general, mass loss rates derived from UV lines are lower than mass loss rates obtained from H${\alpha}$. The UV mass loss rates are even lower than the single line driving limit in the latest type dwarfs. These issues are discussed in the context of the structure of massive stars winds.
 The evolutionary and spectroscopic masses are in agreement above 25 \msun\ but the uncertainties are large. Below this threshold, the few late--type O stars studied here indicate that the mass discrepancy still seems to hold. }
    {}


   \maketitle


\section{Introduction}
\label{s_intro}

Massive stars play key roles in various fields of astrophysics due to their strong ionizing fluxes, powerful winds and extreme stellar properties. They have been recognized as the precursors of long--soft gamma--ray bursts. Population III stars are thought to be made of very massive stars with exceptionally hard UV luminosity. As they live rapidly, they are excellent tracers of star formation. But they also trigger the formation of new generations of stars through their feedback effects. The heavy elements produced in their cores are injected in the interstellar medium by stellar winds and supernovae ejecta, contributing to the chemical evolution of galaxies. Constraining their properties and understanding quantitatively their evolution and feedback effects is thus of interest for fields well beyond pure stellar physics. 

In addition to their mass, the evolution of massive stars is governed by two main parameters: mass loss \citep{cm86} and rotation \citep{mm00}. The accurate determination of mass--loss rates relies mainly on two types of methods. First, the infrared-millimeter-radio excess emission is measured and converted to mass--loss rate \citep{wb75,barlow77,lei91}. This method suffers from little approximations but requires stars to be relatively nearby to be detected with the current generation of instruments. Second, spectroscopic features sensitive to the wind extension (such as P--Cygni profiles or emission lines) are fitted by atmosphere models \citep{puls96,hm99}. This method allows the analysis of larger samples of massive stars, but relies on the physics and approximations included in atmosphere models. The main uncertainty in mass--loss rate determinations (which affects both types of methods) is the non--homogeneity of massive stars winds (clumping). Direct \citep{eversb98} and indirect \citep[e.g.][]{hil91,jc05} evidence exists for the presence of inhomogeneities in O and Wolf--Rayet stars winds. Although theoretically expected \citep{mg79,or84,or85,runacres02}, clumping is currently not fully understood and characterized, although work is in progress \citep[e.g.][]{puls06,sund10}. This is a severe problem for stellar evolution since mass loss is a key parameter. Hence, having as many stars as possible studied quantitatively is desirable in order to understand their outflows. In this context, the late O-type main--sequence stars represent a puzzle of their own, since they display very weak winds with mass--loss rates down to a few $10^{-10}$ \myr\ \citep{n81,ww05,wag09}. The origin of such a weakness is not understood.    

Rotation has several effects on the appearance and evolution of massive stars. In addition to shaping the surface of the star and modifying the surface temperature (poles being hotter than the equator), it triggers internal mixing processes (mainly meridional circulation and shear turbulence) which affect the transport of angular momentum and chemical species. Consequently, the timescales (such as main--sequence lifetime) and surface abundances depend on the rotation rate \citep{mm00}. \citet{hunter08,hunter09} have questioned the role of rotational mixing in the surface chemical enrichment of B stars in the Magellanic Clouds and the Galaxy. They found both slowly rotating N--rich and rapidly rotating non enriched stars that, according to the authors, cannot be accounted for by single star evolution with rotation. However, \citet{maeder09} argued that when care is taken in disentangling the various parameters on which the surface nitrogen content depends, 80\% of the population of B stars is explained by normally rotating single stars. Of particular importance is the age of the population: it is crucial to build samples of stars of the same age to avoid too large a spread in nitrogen surface abundances. So far, the relation between chemical enrichment and rotation has not been looked at in more massive O stars. Since they evolve more rapidly and produce a larger enrichment than B stars, they are important objects to analyze.

The region of the Rosette nebula is especially interesting in this context. It harbors the ionizing cluster NGC2244 at the core of the nebula itself, as well as several more widely spread massive stars in the MonOB2 association. NGC2244 contains seven O stars covering the entire range of spectral types, from O4 to O9. All of them have main--sequence luminosity classes (see Table\ \ref{tab_obs}). The stars of MonOB2 have similar spectral types. Hence, this region contains a homogeneous sample of O stars with different masses and luminosities but most likely of the same age. Binaries are also present \citep{mahy09}. As such, this sample is well suited to investigate the effects of rotation on massive stars surface abundances. The Rosette nebula is also famous for the numerous young stellar objects and embedded clusters in the so-called Rosette molecular cloud \citep{pl97,dif10}. Triggered star formation is thought to be at work in this region, due to the ionizing radiation and powerful winds of the NGC2244 O stars. Recent results obtained by the {\it Herschel Space Observatory} reveal the presence of younger objects further away from the HII region center, as well as temperature gradients \citep{dif10,motte10,schneider10}. Constraining the mechanical energy release and ionizing flux of the massive star population is thus important in order to quantitatively study the process of star formation triggering.   

In this paper, we analyze nine O stars and one supposed B star in the Rosette nebula region. Seven are located in NGC2244, three in MonOB2. Our study encompasses six out of the seven O stars powering the Rosette HII region\footnote{Only the O9V star HD258691 is not included due to the lack of spectroscopic data.}. Two confirmed binaries (HD46149 and HD46573) and one binary candidate (HD46150) are included. This work complements and extends the analysis of HD48099 \citep{mahy10}, also a binary member of the MonOB2 association. In Sect.\ \ref{s_obs} we present the observations of the targets. Their analysis by means of atmosphere models is described in Sect.\ \ref{s_models}. The results are given in Sect.\ \ref{s_results} and discussed in Sect.\ \ref{s_disc}. Finally, our conclusions are drawn in Sect.\ \ref{s_conc}.


\section{Observations and data reduction}
\label{s_obs}

\begin{table*}
\begin{center}
\caption{Sample stars and photometry.} \label{tab_obs}
\begin{tabular}{llcccccccccccc}
\hline
Source          & ST & U & B & V & J & H & K & A$_{\rm V}$& M$_V$ \\
                &    &   &   &   &   &   &  & & \\
\hline   
                &    &   &   &   NGC2244   &     &      &  & \\
\hline
HD 46223   & O4V((f))   &   6.712 &  7.484 &  7.266 &  6.742 &  6.684 &  6.693 & 1.54 & $-$5.22 \\
HD 46150   & O5V((f))z  &   6.043 &  6.875 &  6.745 &  6.446 &  6.451 &  6.453 & 1.27 & $-$5.48 \\
HD 46485   & O7Vn       &   7.876 &  8.563 &  8.243 &  7.565 &  7.492 &  7.463 & 1.86 & $-$4.57 \\
HD 46056   & O8Vn       &   7.705 &  8.445 &  8.245 &  7.837 &  7.816 &  7.837 & 1.49 & $-$4.20 \\
HD 46149-1 & O8V        &   6.982 &  7.774 &  7.602 &  7.245 &  7.232 &  7.268 & 1.40 & $-$4.13 \\
HD 46149-2 & O8.5--9V   &      -- &     -- &     -- &     -- &     -- &     -- & 1.40 & $-$3.84 \\
HD 46202   & O9.5V      &   7.623 &  8.360 &  8.183 &  7.784 &  7.760 &  7.737 & 1.42 & $-$4.19 \\
\hline   
                &    &   &   &   Mon OB2   &     &      &  & \\
\hline
HD 46573   & O7V((f))z  &   7.613 &  8.273 &  7.933 &  7.197 &  7.148 &  7.145 & 1.92 & $-$4.94 \\
HD 48279   & ON8.5V     &   7.254 &  8.047 &  7.910 &  7.650 &  7.681 &  7.710 & 1.29 & $-$4.33 \\
HD 46966   & O8.5IV     &   5.910 &  6.827 &  6.877 &  6.919 &  6.951 &  7.035 & 0.71 & $-$4.78 \\
HD 48099-1 & O5.5V((f)) &   5.387 &  6.323 &  6.366 &  6.445 &  6.490 &  6.512 & 0.73 & $-$5.39 \\
HD 48099-2 & O9V        &      -- &     -- &     -- &     -- &     -- &     -- & 0.73 & $-$3.89 \\
\hline
\end{tabular}
\tablefoot{Spectral types are from \citet{sota11} and \citet{mahy09}. Magnitudes are from \citet{gos}. The extinction A$_{\rm V}$ is 3.1 $\times$ E(B$-$V). The absolute magnitude assumes a distance of 1.55 kpc. For binary stars, we give the observed magnitudes of the system on the row of the primary star. For
binary systems, the absolute V magnitudes of each component are computed
assuming the V--band flux ratio typical for stars of the same spectral type (see Sect.\ \ref{s_results}).}
\end{center}
\end{table*}

Table\ \ref{tab_obs} summarizes the main observational properties of
our sample stars. Photometry is taken from the Galactic O star Catalog
of \citet{gos}. We have adopted a distance of 1.55$\pm$0.15 kpc for the Mon OB2
association and NGC2244. Photometric distances range between 1.4 and 1.7 kpc
according to the summary of \citet{hens00}. We simply adopt the
average of these values in the present work \citep[see also][]{mahy09}.

Spectroscopic data were obtained at the Observatoire de Haute-Provence
(OHP, France) with the Aur\'elie spectrograph mounted on the 1.52m
telescope. Several observing runs from 2006 to 2008 allowed us to
collect spectra in two different wavelength domains: [4450--4900] and
[5500-5920]\AA. In the adopted configuration, the Aur\'elie
spectrograph offers a resolving power of about $R$~=~$8\,000$ in the
blue domain whilst $R$~=~$10\,000$ is reached in the red band. The
exposure times, necessary to reach a signal-to-noise ratio higher than
150, were typically 10 to 45 min. We also retrieved 13 spectra from
the Elodie ($R$~=~$42\,000$) archives taken from November 1999 to
November 2005 with the 1.93m telescope at the OHP. The exposure times
of these data ranged from 10 min to 1h.

In April 2007 and March 2008, another set of spectra was obtained at
Observatorio Astron\'omico Nacional of San Pedro M\`artir (SPM) in
Mexico with the Espresso spectrograph mounted on the 2.10m
telescope ($R$ = $27\,000$). This \'echelle spectrograph covers the wavelength domain
[3780--6950]\AA\ with 27 orders. We used a typical exposure time
from 5 to 15 min. Several consecutive spectra of a given night were
added to improve the signal-to-noise ratio of our data.

We have also retrieved 11 spectra from the ESO archives (PI: Casassus,
run 076.C-0431(A) and PI: Lo Curto, run 076.C-0164(A)) taken with the
Fiber-fed Extended Range Optical Spectrograph (FEROS) mounted on the
ESO/MPG 2.20m telescope at La Silla (Chile). The wavelength region
covered by this instrument is [3550--9200]\AA\ and the spectral
resolution is about $R$~=~$48\,000$. The data reduction was done
with an improved version of the FEROS pipeline as described in
\citet{sana09}.

Finally, to model the winds of the O-type stars with more accuracy, we
have retrieved a total of 40~high-resolution {\it IUE} spectra (SWP
with a dispersion of 0.2\AA). Such data are available for all targets
of our sample, except HD\,46573. These data were taken between
September 1978 and February 1987 with exposure times between 539 and
18000 s. These spectra cover the [1200--1800]\AA~spectral band, which
is particularly interesting for the UV~P-Cygni profiles
($\ion{C}{iv} 1548-1550$ and $\ion{N}{v} 1240$).

A full description of the optical instruments used as well as the
reduction procedure was given in
\citet{mahy09}. Tables\ \ref{tab_data_obs} and
  \ref{tab_data_iue} present the main characteristics of the
  observational data. For the single stars, we have selected the
  spectra of \citet{mahy09} with the widest wavelength coverage,
  highest spectral resolution and highest signal--to--noise ratio.

\begin{table*}
\caption{Journal of optical observations.}              
\label{tab_data_obs}      
\centering           
\begin{tabular}{llcccccc}
\hline\hline         
Star & Instrument & Resolution & Spectral range & S/N & Obs Date & HJD & Exposure time\\
     &            &            &   [\AA]        &     &          & [d] &  [s] \\
\hline                                   
HD\,46056 & ESPRESSO & 18000 & $[3950-6700]$ & 269 & 20 Mar 2008 & 2454545.6733 & 2700\\
HD\,46149 & \multicolumn{7}{c}{disentangled spectrum -- see data in \citet{mahy09}}\\
HD\,46150 & FEROS    & 48000 & $[4000-7100]$ & 310 & 06 Feb 2006 & 2453772.5895 & 3600\\
HD\,46202 & FEROS    & 48000 & $[4000-7100]$ & 154 & 05 Jan 2006 & 2453740.6191 & 600\\
HD\,46223 & FEROS    & 48000 & $[4000-7100]$ & 194 & 08 Feb 2006 & 2453774.5460 & 3600\\
HD\,46485 & FEROS    & 48000 & $[4000-7100]$ & 166 & 05 Jan 2006 & 2453740.6290 & 600\\
HD\,46573 & FEROS    & 48000 & $[4000-7100]$ & 268 & 07 Feb 2006 & 2453773.5904 & 3600\\
HD\,46966 & FEROS    & 48000 & $[4000-7100]$ & 200 & 03 Jan 2006 & 2453738.7201 & 600\\
HD\,48099 & \multicolumn{7}{c}{disentangled spectrum - see \citet{mahy10}}\\
HD\,48279 & ELODIE   & 42000 & $[4000-6800]$ & 214 & 25 Jan 2000 & 2451569.5028 & 2700\\
\hline                                         
\end{tabular}
\end{table*}

\begin{table*}
\caption{List of IUE UV spectra.}              
\label{tab_data_iue}      
\centering           
\begin{tabular}{llccc}
\hline\hline         
Star & data ID       & Obs. Date & HJD & Exposure time\\
     &               &           & [d] &  [s] \\
\hline                                   
HD\,46056 & SWP06949 & 22 Oct 1979 & 2444168.92297 & 5100\\
HD\,46149 & SWP28151 & 11 Apr 1986 & 2446532.40344 & 2640\\
HD\,46150 & SWP10758 & 05 Dec 1980 & 2444578.86527 & 1200\\
HD\,46202 & SWP30299 & 12 Feb 1987 & 2446839.37549 & 5100\\
HD\,46223 & SWP10757 & 05 Dec 1980 & 2444578.81523 & 1980\\
HD\,46485 & SWP28196 & 19 Apr 1986 & 2446539.98248 & 10800\\
HD\,46966 & SWP24196 & 18 Oct 1984 & 2445991.75211 & 540\\
          & SWP27918 & 15 Mar 1986 & 2446505.35266 & 540\\
HD\,48099 & \multicolumn{4}{c}{disentangled spectrum - see \citet{mahy10}}\\
HD\,48279 & SWP06504 & 14 Sep 1979 & 2444130.71788 & 7200\\
\hline                                         
\end{tabular}
\tablefoot{IUE UV spectra retrieved from the archive and used for the analysis. The wavelength range is 1150--1975 \AA\ and the spectral resolution 0.2 \AA\ ($R$=7500 at 1500 \AA) for all spectra. For HD~48279, we used the average of the two spectra listed in the table since they were obtained in the same mode (high aperture), with the same exposure time and do not show any variability. For the other stars, we chose the spectra with the highest exposure time in the high aperture mode.}
\end{table*}


\section{Modelling}
\label{s_models}

We have used the code CMFGEN \citep{hm98} for the quantitative analysis of the optical and UV spectra. CMFGEN provides non--LTE atmosphere models including winds and line--blanketing. An exhaustive description can be found in \citet{hm98} and a summary of the main characteristics is given in \citet{hil03}. CMFGEN needs as input an estimate of the hydrodynamical structure that we constructed from TLUSTY models \citep[taken from the OSTAR2002 grid of][]{lh03} connected to a $\beta$ velocity law of the form $v = v_{\infty} (1-R/r)^{\beta}$ where \vinf\ is the wind terminal velocity. We adopted $\beta$ = 0.8 since this is the typical value for O dwarfs \citep[e.g.][]{repolust04}. Once the model was reasonably converged, the radiative acceleration was computed and used to iterate on the initial hydrodynamical structure in the inner atmosphere. The radiation field and level populations were then converged once again. Two to five of these global hydrodynamical iterations were performed before the atmosphere model finally converged. 

Our final models include H, He, C, N, O, Ne, Mg, Si, S, Fe with the solar composition of \citet{gre07} unless otherwise stated. The super--level approach is used to reduce the amount of memory requirements. On average, we include 1600 super levels for a total of 8000 levels. For the formal solution of the radiative transfer equation leading to the emergent spectrum, a microturbulent velocity varying linearly (with velocity) from 10 \kms\ to 0.1 $\times$ \vinf\ was used.

We included X--ray emission in the wind since this can affect the ionization balance and the strength of key UV diagnostic lines. In practice, we adopted a temperature of 3 million degrees and adjusted the flux level so that the X--ray flux effectively coming out of the atmosphere matches the observed $L_{X}$/$L_{\mathrm{bol}}$ ratio. This ratio was computed from the observed X--ray fluxes of \citet{wang08} and our derived luminosities. In case no X--ray flux was detected, we simply adopted the canonical value $L_{X}$/$L_{\mathrm{bol}}$ = 10$^{-7}$ \citep{sana06,naze09}.

In practice we proceeded as follows to derive the stellar and wind parameters:

\begin{itemize}

\item[$\bullet$] \textit{Effective temperature}: we used the classical ratio of \ion{He}{i} to \ion{He}{ii} lines to constrain \teff. The main indicators were \hei\ and \heii. Additional diagnostics are \ion{He}{i} 4026 (blended with \ion{He}{ii} 4026), \ion{He}{i} 4388, \ion{He}{i} 4713, \ion{He}{i} 4921, \ion{He}{i} 5876, \ion{He}{ii} 4200, \ion{He}{ii} 5412. The typical uncertainty on the \teff\ determination is 1000 K.

\item[$\bullet$] \textit{Gravity}: The wings of the Balmer lines H$\beta$, H$\gamma$ and H$\delta$ are the main \logg\ indicators. An accuracy of about 0.1 dex on \logg\ is achieved.

\item[$\bullet$] \textit{Luminosity}: We fitted the flux calibrated IUE spectra and UBVJHK fluxes to derive both the luminosity and the extinction. For this, we adopted a distance of 1.55$\pm0.15$ kpc \citep[see discussion in][]{mahy09}. The Galactic extinction law of \citet{seaton79} and \citet{howarth83} was used. We adjusted the stellar luminosity and E(B$-$V) to reproduce the UV-optical-infrared SED. In those cases where no UV spectra were available, we simply derived the extinction from E(B$-$V) and adjusted the luminosity to reproduce the V magnitude.  

\item[$\bullet$] \textit{Wind terminal velocity}: the blueward extension of P--Cygni lines observed in the UV provides \vinf\ with an accuracy of 100 \kms. 

\item[$\bullet$] \textit{Mass--loss rate}: We used two diagnostic lines to constrain the mass loss rate: the UV P--Cygni profiles and H${\alpha}$. In principle, a single value of \mdot\ should allow a good fit of both types of lines. However in practice, several recent studies have shown that this was not the case. In order to be as complete as possible, we thus provide both values: the ``UV mass loss rate'' ($\dot{M}_{\mathrm{UV}}$, given in Table\ \ref{tab_param}) and the ``H${\alpha}$ mass loss rates'' ($\dot{M}_{\mathrm{H\alpha}}$, given in Table\ \ref{tab_lc2}).

\end{itemize}

The projected rotational velocities were derived by direct comparison of our synthetic spectra to observed line profiles for fast rotating stars (i.e., those with \vsini\ $\geq$ 150 \kms). The rotationally broadened synthetic line profiles usually provided a good match to the observed profile. In the case of moderately to slowly rotating objects, some amount of macroturbulence had to be introduced to correctly reproduce the line profiles of several features (e.g., \ion{He}{i} 4713, \ion{C}{iv} 5812, \ion{He}{i} 5876). The need for extra broadening is well documented \citep{howarth97,ryans02,howarth07,nieva07,martins10,fraser10} but its origin is unclear. \citet{aerts09} recently suggested that non--radial pulsations could trigger large--scale motions (hence macroturbulence), but this needs to be confirmed by a study covering a wider parameter space. A recent study by \citet{sergio10} showed that the amplitude of macroturbulence was correlated to the amplitude of line profile variability in a sample of OB supergiants. In practice, we introduced macroturbulence by convolving our (rotationally broadened) synthetic spectra with a Gaussian profile ($\propto e^{-\frac{v^2}{2 v_{\mathrm mac}^2}}$ where \vmac\ is the macroturbulent velocity), thus mimicking isotropic turbulence. This is obviously a very simple approach, but it significantly improves the quality of the fits. Given the present limitations, we restricted ourselves to rough estimates of the amount of macroturbulence by judging the fit quality by eye. In practice, we used \ion{He}{i} 4713 as the main indicator of macroturbulence since it is present with sufficient SNR in all our sample stars. Secondary indicators were the \ion{C}{iv} doublet at 5801--5812 \AA\ and \ion{He}{i} 5876. \vsini\ was also obtained from the Fourier transform method \citep{sergio07}. In practice, for the moderately to slowly rotating stars, we obtained upper limits from the absence of zero and the position of the noise level in the FT.

We also determined the nitrogen content of our sample stars. We relied mainly on the \ion{N}{iii} lines between 4500 and 4520 \AA. They are present in absorption in all the stars. They are not affected by winds and are strong enough for abundance determination. The uncertainties are of the order of 50\%. They were estimated from the comparison of the selected \ion{N}{iii} lines to models with various N content. The errors do not take into account any systematics related to atomic data. 

When possible, the degree of inhomogeneities of the winds of our sample stars was determined. Clumping is implemented in CMFGEN by means of a volume filling factor $f$ following the law $f = f_{\infty} + (1-f_{\infty}) e^{-v/v_{cl}}$ where $f_{\infty}$ is the maximum clumping factor at the top of the atmosphere and $v_{cl}$ a parameter indicating the position where the wind starts to be significantly clumped. As shown by \citet{jc05}, \ion{O}{v} 1371 and \ion{N}{iv} 1720 are two UV features especially sensitive to wind inhomogeneities in early O stars. We have used these lines to constrain $f_{\infty}$ in the earliest O stars of our sample. In mid to late O stars, no UV clumping diagnostic has been identified. This is confirmed by our study. For the O7--O9 stars of our sample, we computed models with several values of $f_{\infty}$ (scaling \mdot\ so that $\dot{M}/\sqrt{f}$ is constant) and found no difference in the UV wind sensitive lines. Hence, we decided to adopt $f_{\infty}=1$ (i.e. homogeneous model). Our mass--loss rates for those objects should thus be regarded as upper limits.


\section{Results}
\label{s_results}

\begin{figure*}[t]
\centering
\includegraphics[width=19cm]{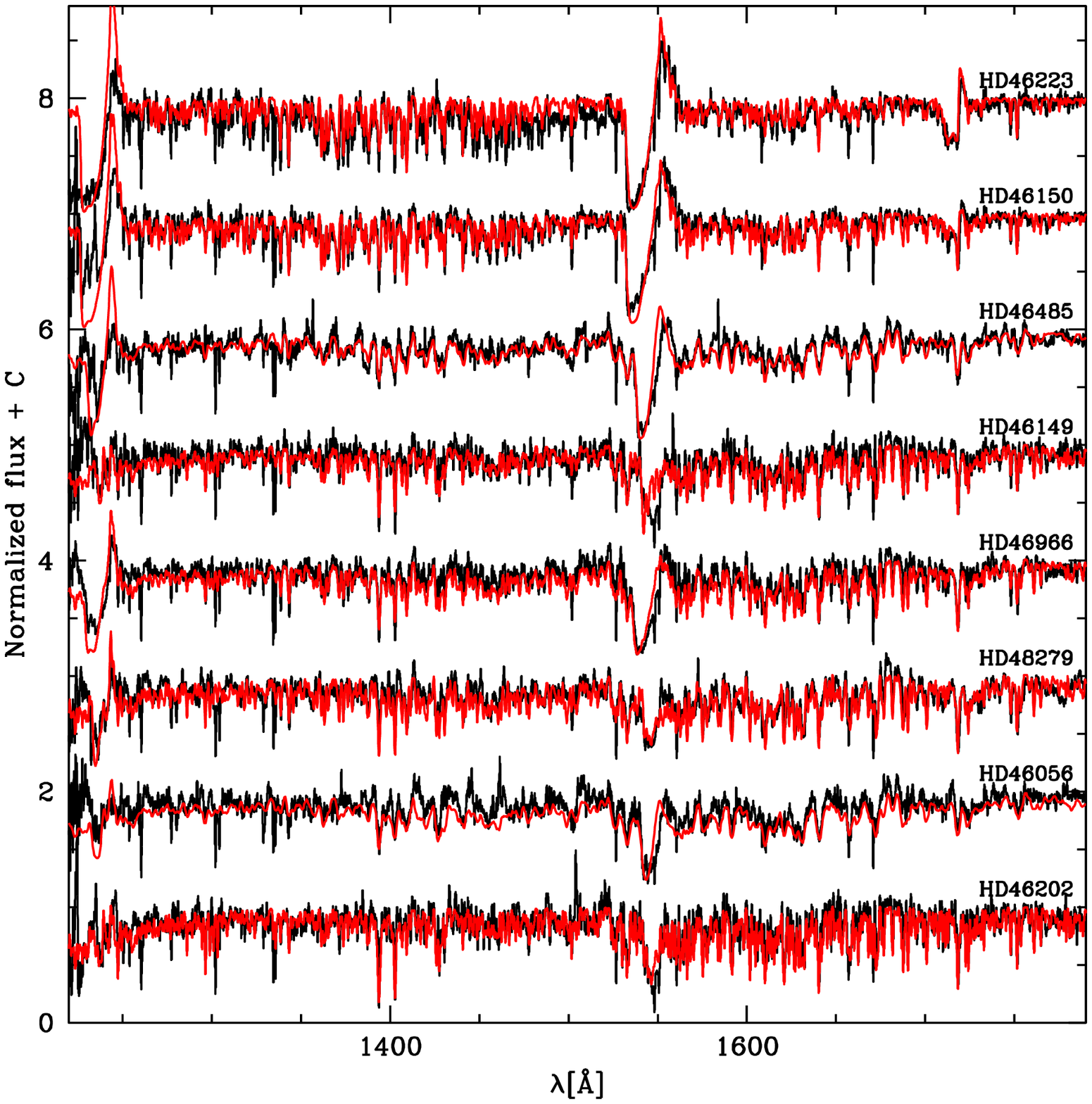}
\caption{Best CMFGEN fits (red) of the UV spectra (black) of the sample stars.} \label{fit_uv}
\end{figure*}

The derived stellar and wind parameters of the sample stars are gathered in Table \ref{tab_param}. The values of the H$\alpha$ mass loss rates are given in Table \ref{tab_lc2}. We show the best optical fits for HD~46223 and HD~48279 in Figs.\ \ref{fit_opt_1}. The optical fits for the other stars are gathered in Appendix A. The fits to the UV spectra for all stars are shown in Fig.\ \ref{fit_uv}. Below, we briefly comment on each star.

\begin{sidewaystable*}
\vspace{16cm}
\begin{center}
\caption{Derived stellar properties of the sample stars.} \label{tab_param}
\begin{tabular}{llcccccccccccccccc}
\hline
Source          & ST & \teff\ & \lL\ & log(g) & log($\dot{M}_{\mathrm{UV}}$) & $f_{\infty}$ &\vinf & \vsini\ & \vmac\ & M$_{evol}$ & M$_{spec}$ & N/H & log(Q$_{0}$)\\
                &    & [kK]   &      &        &            &     &[\kms]& [\kms]  & [\kms] & [\msun]   & [\msun]   & [$\times 10^{4}$] & \\
\hline  
                &    &        &      &        &  NGC2244   &     &      &         &        &           &      & \\
\hline
HD 46223       & O4V((f))   &  43.0 & 5.60$\pm$0.11 & 4.01 & $-$7.17   & 0.1 & 2800 & 100     & 32 & $52.1^{+6.2}_{-5.9}$ & $48.3\pm 19.3$  & 7.0$\pm$2.0 & 49.40 \\
HD 46150       & O5V((f))z  &  42.0 & $<$5.65       & 4.01 & $-$7.30   & 0.1 & 2800 & 100     & 37 & $<53.6$           & $59.5\pm 23.7$  & 3.0$\pm$2.0 & 49.36 \\
HD 46485       & O7Vn       &  36.0 & 5.05$\pm$0.11 & 3.85 & $-$7.80   & 1.0 & 1850 & 300     & -- & $28.6^{+3.3}_{-4.3}$ & $19.5\pm 7.9$   & $\leq$1.2 & 48.62 \\
HD 46056       & O8Vn       &  34.5 & 4.85$\pm$0.12 & 3.89 & $-$8.50   & 1.0 & 1500 & 330     & -- & $23.1^{+2.4}_{-1.9}$ & $15.8\pm 6.7$   & $\leq$0.6 & 48.32 \\
HD 46149-1     & O8V        &  37.0 & 4.90$\pm$0.12 & 4.25 & $<-$9.0 &     & 1300 & $\sim$0 & 24 & $25.9^{+3.1}_{-3.5}$ & $30.9\pm 21.6$  & 0.8$\pm$0.5  & 48.42 \\
HD 46149-2     & O8.5--9V   &  35.0 & 4.65$\pm$0.16 & 4.01 & --      & --  & --   & 100     & 27 & $20.6^{+3.4}_{-4.1}$ & $12.5\pm 10.1$    & 0.8$\pm$0.5  & 48.10 \\
HD 46202       & O9.5V      &  33.5 & 4.85$\pm$0.12 & 4.10 & $-$9.0    & 1.0 & 1200 & 20      & 17 & $22.7^{+2.0}_{-2.4}$ & $29.0\pm 12.4$  & 1.0$\pm$0.5  & 48.19 \\
\hline  
               &    &        &      &        &  Mon OB2   &     &      &         &        &           &      \\
\hline
HD 46573       & O7V((f))z  &  36.5 & $<$5.20       & 3.75 & --      & --  & --   & 20      & 43 & $<32.5$           & $20.6\pm 8.3$  & 3.0$\pm$1.0 & 48.84 \\
HD 48279       & ON8.5V     &  34.5 & 4.95$\pm$0.11 & 3.77 & $-$8.8    & 1.0 & 1300 & 125     & 22 & $24.4^{+2.5}_{-1.9}$ & $15.2\pm 6.1$  & 5.0$\pm$3.0 & 48.47 \\
HD 46966       & O8.5IV     &  35.0 & 5.20$\pm$0.11 & 3.75 & $-$8.0    & 1.0 & 2300 & 50      & 27 & $30.9^{+3.8}_{-4.1}$ & $24.5\pm 9.9$  & 1.2$\pm$0.5 & 48.70 \\
HD 48099-1\tablefootmark{a} & O5.5V((f)) &  44.0 & 5.65$\pm$0.07 & 4.50 & $-$7.60   & 0.01& 2800 & 91      & 0  & $55.3^{+7.3}_{-5.0}$ & $155.0\pm 98.7$ & 5.0$\pm$2.5 & 49.36 \\
HD 48099-2\tablefootmark{a} & O9V        &  32.0  & 4.60$\pm$0.06 & 3.51 & --     & --  & --   & 51      & 0  & $18.5^{+2.2}_{-1.7}$ & $5.0\pm 3.5$    & 0.6$\pm$0.3 & 47.93 \\

\hline
\end{tabular}
\end{center}
\tablefoot{The columns give: name, spectral type, effective temperature (uncertainty 1000K except for HD~46149--1 / HD~48099--1, 2000K and HD~46149--2 / HD~48099--2, 3000K), luminosity, effective gravity (uncertainty 0.1 dex, except for HD~46149 and HD~48099, 0.2 dex), UV mass--loss rate, terminal velocity, projected rotational velocity, macroturbulent velocity, evolutionary mass, spectroscopic mass, nitrogen content, ionizing flux. Spectroscopic masses are computed from the true gravity (i.e. \logg\ corrected from the effect of centrifugal force).}
\tablefoottext{a}{from \citet{mahy10}, except that the error bars on \teff\ and \logg\ are similar to HD~46149 (see text for discussion).}
\end{sidewaystable*}

\begin{figure*}[!ht]
     \centering
     \subfigure[HD 46223]{
          \includegraphics[width=15cm,height=11cm]{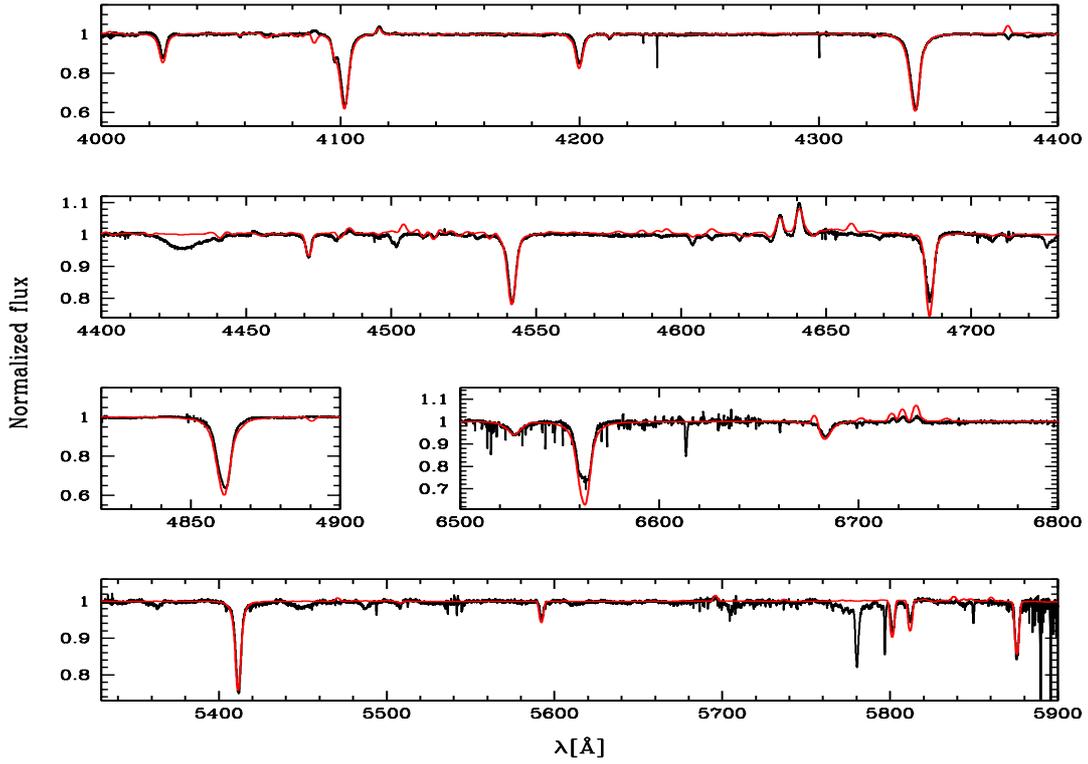}}\\
     \subfigure[HD 48279]{
          \includegraphics[width=15cm,height=11cm]{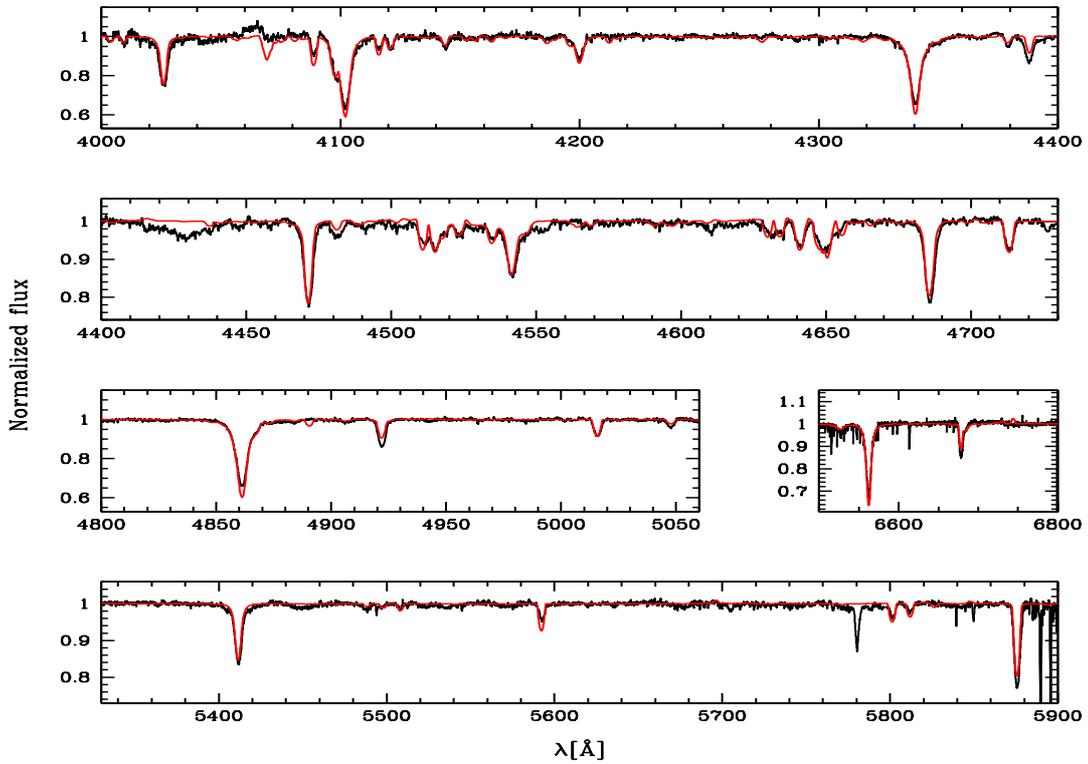}}
     \caption{Best CMFGEN fits (red) of the optical spectra (black) of stars HD46223 (top), HD48279 (bottom). The models have the mass loss rate best reproducing the UV features. The unfitted feature around 4425 \AA\ is a DIB.}
     \label{fit_opt_1}
\end{figure*}

\textit{HD~46223}: a clumped wind ($f$=0.1) is required to fit \ion{O}{v} 1371. The nitrogen content is difficult to estimate given the weakness of the \ion{N}{iii} 4500--4520 features. A value of $N/H$=7.0 $\times\ 10^{-4}$ is preferred. It also provides a good fit to the \ion{N}{iv} 1720 feature. The parameters are similar to those of \citet{ww05} within the uncertainties. The better quality and wider wavelength coverage of the present optical spectra explains the refined effective temperature estimate.

\textit{HD~46150}: this is a binary candidate. We thus derive upper limits on the luminosity and mass loss rate. A clumping factor of 0.1 is necessary to correctly reproduce the \ion{O}{v} 1371 \AA\ line profile. Models with higher (lower) clumping factor yield a too strong (weak) absorption in the blue part of the profile. See also Sect.\ \ref{s_wind} for further discussion. 

\textit{HD~46485}: this is the second fastest rotator of our sample, with \vsini\ = 300 \kms. The fit of the optical and UV spectra is of excellent quality. The only notable exceptions are the \ion{C}{III} line complex at 4640 \AA, \ion{C}{iv} 5801--5812 \AA\ and \ion{He}{i} 5876 \AA. The \ion{C}{iii} 4650 line complex is slightly too weak in our best fit model. However, we found out that this transition is sensitive to detailed line--blanketing effects. The lower level of these transitions is connected to the \ion{C}{iii} ground level by a transition at 538\AA. This transition is blended with \ion{Fe}{iv} transitions. The interaction between these UV lines thus affects the strength of the \ion{C}{iii} 4650 lines, in a manner similar to that described by \citet{paco06} for the \ion{He}{i} singlet lines.

\textit{HD~46056}: this is the fastest rotator of our sample and no macroturbulence is necessary to reproduce the line profiles. The fit is good with the exception of the \ion{C}{iii} 4650 line complex which is underpredicted. The \ion{C}{iii} 4650 lines suffer from the same problem as for HD~46485.

\textit{HD~46149}: this is a binary \citep{mahy09}. The spectra have been disentangled using the method described in \citet{mahy10}. They have subsequently been used to classify the two components of the system. The classical diagnostic log~$\frac{EW(HeI 4471)}{EW(HeII 4542)}$ lead to a spectral type O8 for the primary and O8.5--O9 for the secondary. Similarly, a luminosity class V was assigned based on the values of log $\frac{EW(SiIV 4088)}{EW(HeI 4143)}$ \citep{ca71}. These classifications are slightly different from those derived by Mahy et al (2009) although for the primary it remains in the error range. However, such a discrepancy for  the secondary could come from the disentangling process, making the disentangled spectra uncertain. Indeed, they were computed from a sample of observed spectra taken essentially at similar phases where the separation between the radial velocities of both components ($RV_S - RV_P$) are of about $-30$, $0$ and $+140$~\kms. Moreover, the orientation and the large eccentricity of the orbit imply an asymmetric excursion in radial velocity which could also affect the accuracy of the broadest lines. We thus caution that the disentangling process does not provide perfect results and should be refined in the future when more data taken at different phases will be obtained. This is especially true for the wings of broad lines which need to be correctly sampled and fully separated in the observed spectrum (at maximum separation) to be reconstructed correctly (see below -- HD~48099). Consequently, the gravity and spectroscopic masses in Table \ref{tab_param} are highly uncertain. The effective temperature and nitrogen content have also larger error bars than the other stars of our sample: 2000 K (3000 K) on \teff\ for the primary (secondary) and $\sim$70\% on N/H. In addition, given the poor quality of the spectrum of the secondary, we adopted \logg\ = 4.0. Using the calibrations of \citet{mp06}, we find that the V--band flux ratio between a O8V and a O9V star is $\approx$ 1.3. Optical photometry of HD~46149 indicates E(B$-$V)= 0.45, leading to $A_{\rm V}$ = 1.40. For the adopted distance, the absolute V--band magnitude of the system is thus $M_V$ = $-$4.75. Given the V--band flux ratio of the two components, we have $M_{\mathrm V}^{\rm primary}$ = $-$4.13 and $M_{\mathrm V}^{\rm secondary}$ = $-$3.84. 

\textit{HD~46202}: the parameters derived for HD~46202 are very similar to those of \citet{ww05}. The fit to the optical spectrum is of excellent quality. A combination of a low rotation rate (20 \kms) and a macroturbulent velocity of 17 \kms\ correctly reproduces the photospheric line profiles. See also Sect.\ \ref{s_wind} for further discussion.

\textit{HD~46573}: this is a binary candidate. No UV spectra are available for that star, so that our analysis is restricted to the optical range. A very good fit is achieved. Only the core of the Balmer lines is too deep in our synthetic spectra. The line profiles show that macroturbulence is important. We found that a combination of \vsini\ in the range 40--60 \kms\ with a macroturbulent velocity of 40 to 50 \kms\ leads to much better fits compared to purely rotationally broadened profiles. 

\textit{HD~48279}: our synthetic spectrum provides a good fit of most optical and UV lines. The singlet lines \ion{He}{i} 4388 \AA\ and \ion{He}{i} 4921 \AA\ are too weak in our spectrum. They are known to be sensitive to details of line blanketing \citep{paco06}. 

\textit{HD~46966}: the fit of the optical and UV spectra is one of the best of the present work. The only problem is observed in \ion{He}{ii} 4686 \AA\ for which our synthetic profile is too deep. A projected rotational velocity of the order 40--50 \kms\ combined with a macroturbulent velocity $\sim$ 30--40 \kms\ provided better fits than models with only rotationally broadened profiles.

\textit{HD~48099}: this star is a binary and has been studied by
\citet{mahy10}. We have not re-analyzed the two components and we have
simply adopted the parameters derived by Mahy et al. However, we add a
few words of caution. The spectra used in the disentangling process of
this binary system sample the full orbital cycle and hence do not
suffer from the same limitations as in the case of HD46149. But whilst
we expect narrow spectral features to be reliably reconstructed in the
disentangled spectra of both the primary and secondary star, the
reconstruction of the broad features is more uncertain. Indeed, given
the rather low orbital inclination of HD48099 \citep{mahy10}, the
maximum radial velocity excursions of the stars remain small
($RV^{max}_P - RV^{max}_S \sim 155$ \kms\ -- $RV^{max}_P$ and
$RV^{max}_S$ being the maximum radial velocities of the primary and
secondary components) compared to the possible widths of broad Balmer
lines ($\pm 950$\kms\ for H$\gamma$). This poor sampling of the
broadest lines can lead to a poor reconstruction of the wings of the
Balmer lines especially for the fainter secondary component. Hence,
the gravity of the latter is very likely underestimated. There is
unfortunately no way to remedy to this situation and as a result, the
spectroscopic masses for this system are unreliable as becomes clear
for instance when comparing the spectroscopic mass ratio from Table 4
($M_1/M_2 \sim$ 30) with the dynamical mass ratio ($M_1/M_2 = 1.77$)
from the orbital solution of \citet{mahy10}. In view of this discussion, we chose to adopt the same error bars on \teff\ and \logg\ as for HD~46149 (i.e. 2000 K on \teff\ -- 3000 K for the secondary-- and 0.2 dex on \logg). The error bars are thus larger than in \citet{mahy10}. They are also realistic, since reducing \logg\ by 0.25 dex, we find that we need an effective temperature lower by 1000-2000 K to correctly reproduce the \ion{He}{i} and \ion{He}{ii} lines of the primary.


\section{Discussion}
\label{s_disc}

\subsection{HR diagram and evolutionary status}
\label{s_hrd}

\begin{figure}[!t]
\centering
\includegraphics[width=9cm]{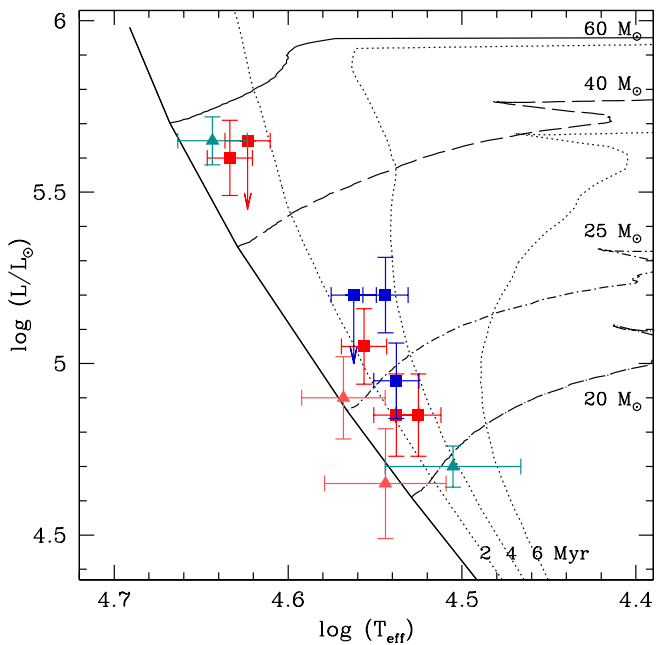}
\caption{HR diagram of NGC2244. The evolutionary tracks and isochrones are from \citet{mm03}. They are computed for an initial rotational velocity of 300 \kms. Triangles are the components of the binaries HD48099 (cyan) and HD46149 (pink). Stars belonging to NGC2244 are shown in red/pink and stars from the Mon~OB2 association are shown in blue/cyan. Arrows indicate upper limits on luminosities for SB1 and binary candidates.} \label{fig_hrd}
\end{figure}

The HR diagram of the stars analyzed in the present study is shown in Fig.\ \ref{fig_hrd}. Red/pink (blue/cyan) colors are used for the stars of NGC2244 (Mon~OB2 respectively). 

The stars of NGC~2244 have an age less than 5 Myr. The most massive stars of the cluster, namely HD~46223 and HD~46150, are less than 2 Myr old. There is no significant age difference between these two stars, contrary to what \citet{wang08} suspected. Whether the trend for the most massive stars being younger than the lower mass O stars is real or an artifact of the parameter determination is unclear. It is known that the stellar parameters of an average early O star indicate a younger age than that of late--type O stars \citep[e.g. Fig.\ 12 of][]{msh05}. An explanation to this trend is that late-type O stars have lower ionizing fluxes and weaker winds than early--type O stars. Consequently, they remain hidden in their parental cloud for a longer time, so that on average late--type O stars appear older. In the present case, since both early and late--type stars are visible at the same time, one needs another explanation. A possibility is that the most massive stars were the last to form. Alternatively, there might be a bias in the determination of the effective temperature of the hottest O stars. It would have to be 2000-3000~K lower in order to lead to an age similar to the late--type O stars. Such a reduction is difficult to explain. Line--blanketing, when included in model atmospheres, leads to a downward revision of effective temperatures by several thousands K. But as shown by \citet{msh02}, C, N, O, Si, S and Fe are the main contributors to opacities in O stars atmospheres. Since our present models include all these elements and additional ones, it is unlikely that we overestimate \teff\ by 2000--3000~K. Since we are using clumped models for the hottest stars and homogeneous winds for the coolest ones, one might wonder whether this affects the temperature determination. However the photospheric lines used to constrain \teff\ are unaffected by clumping since they form deep in the atmosphere where the wind is still homogeneous even in clumped models. Another speculative hypothesis is that the most massive O stars have, on average, larger rotation rates. If so, the use of the same set of evolutionary tracks (i.e., computed with the same initial rotational velocity) for early and late--type O stars is inappropriate. Indeed for faster rotation, the main--sequence lifetime is increased. \citet{mm00} concluded that the use of non--rotating tracks could lead to an underestimate of the actual age of stars by about 25\%. \citet{brott11a} show the effect of rotation on isochrones (see their Fig.\ 7) and confirm fast rotating massive stars stay closer to the zero-age main sequence compared to moderately rotating lower mass stars of the same age. This could partly explain the age difference we observe between early and late O stars. But this relies on the assumption that rotation is faster for larger masses among O stars. So far, there is no detailed study of the evolution of rotational velocities with spectral type among O stars. An average \vsini\ of 129$\pm$82 \kms\ results from the work of \citet{penny96}, consistent with the more recent result of \citet{penny04} -- 131$\pm$93 \kms. \citet{howarth97} found a distribution of \vsini\ peaked at about 100 \kms. Concerning B stars, \citet{abt02} found values of 100--130 \kms for B0--B5V stars. It is thus not clear whether there is a trend of larger rotation rate between B and O stars (and consequently not clear whether such a trend continues within the O class).

The stars in the Mon~OB2 association appear to have an age of about 1--5 Myr, with a preferred value between 2 and 4 Myr. The only exception is the primary of the binary system HD~48099. As already discussed by \citet{mahy10}, several factors can explain the apparent younger age of this star: rapid rotation, mass and angular momentum transfer, inappropriate evolutionary tracks. Rapid rotation was the favored explanation. The present study confirms that the secondary star has an age similar to the other stars of the association, making the primary peculiar. It is thus very likely that the position of the primary in the HR diagram is due to a rapid rotation and that comparison with normal (i.e. moderate rotation) evolutionary tracks is not appropriate.

\subsection{Wind properties}
\label{s_wind}

\begin{table*}
\begin{center}
\caption{Summary of various mass loss rates for our sample stars.} \label{tab_lc2}
\begin{tabular}{lrcccc}
\hline

Star       & ST & log($\dot{M}_{\mathrm{UV}}$) & log($\frac{L}{c^2}$) & log($\dot{M}_{\mathrm{H\alpha}}$) & log($\dot{M}_{Vink}$) \\

\hline
HD 46223   & O4V((f))   & $-$7.17   & $-$7.57 & $-$6.20 & $-$5.75 \\
HD 46150   & O5V((f))z  & $-$7.30   & $-$7.52 & $-$6.40 & $-$5.68 \\
HD 46485   & O7Vn       & $-$7.80   & $-$8.12 & $-$6.45 & $-$6.50 \\
HD 46056   & O8Vn       & $-$8.50   & $-$8.32 &   --    & $-$6.80 \\
HD 46149-1 & O8V        & $<-$9.0   & $-$8.27 &   --    & $-$6.74 \\
HD 46202   & O9.5V      & $-$9.00   & $-$8.32 & $-$7.10 & $-$6.72 \\
HD 46573   & O7V((f))z  & $-$-      & $-$7.97 & $-$6.30 & $-$6.25 \\ 
HD 48279   & ON8.5V     & $-$8.80   & $-$8.22 & $-$6.80 & $-$6.64 \\
HD 46966   & O8.5IV     & $-$8.00   & $-$7.97 & $-$6.40 & $-$6.45 \\
HD 48099-1 & O5.5V((f)) & $-$7.60   & $-$7.52 &   --    & $-$5.66 \\
\hline
\end{tabular}
\end{center}
\tablefoot{The columns give: name; spectral type; derived UV mass loss rate; mass loss rate expected for driving by a single line located at the SED emission peak in the single scattering condition; H${\alpha}$ mass loss rate; theoretical mass loss rates of \citet{vink01}. The filling factor $f_{\infty}$ (see Sect.\ \ref{s_models}) is 0.1 for HD~46223 and HD~46150, 0.01 for HD~48099--1, and 1.0 (i.e. no clumping included) for all other stars.}
\end{table*}

In Table \ref{tab_lc2}, we gather the mass loss rates derived from our quantitative analysis ($\dot{M}_{\mathrm{UV}}$ and $\dot{M}_{\mathrm{H\alpha}}$) as well as the theoretical mass loss rates of \citet{vink01} and the values $L/c^2$. The latter corresponds to the mass loss rates obtained if the driving is due to a single line located at the emission peak of the SED. They are thus a lower limit to the mass loss rate expected by the radiation-driven wind theory. From Table \ref{tab_lc2} we can draw the following conclusions:

\begin{enumerate}
\item All stars have $\dot{M}_{\mathrm{UV}} < \dot{M}_{\mathrm{H\alpha}}$. The difference is between a factor of 10 and a factor of a few 100!

\item The theoretical mass loss rates of \citet{vink01} are systematically larger than the UV mass loss rates. The discrepancy reaches 2 orders of magnitudes in the latest type O stars, as usually seen \citep{ww05,wag09}. The H${\alpha}$ mass loss rates are usually in rather good agreement with the theoretical predictions (when comparisons are made for unclumped \mdot\ for the early O dwarfs).

\item For all late type O dwarfs, the UV mass loss rates are \textit{lower} than the single line driving limit. This is a serious puzzle. If the UV mass loss rates are the correct ones, this implies that line driving is less efficient than thought. It could mean that the line absorbs only at discrete values of velocities in the range 0--\vinf\ (i.e., that absorbing material is present only at specific velocities, see below). For the earliest O dwarfs, the UV mass loss rates are only a factor 3--4 larger than this limit. 

\end{enumerate}

The results of Table \ref{tab_lc2} raise the question of the validity of our mass loss rate determinations. We list a few reasons why they could be biased:

\begin{itemize}
\item
The mass-loss rates are based on a few features, and sometimes based on only a  single line (although the absence of other wind features provides upper limits).
\item
The inferred mass-loss requires that the wind ionization structure derived by CMFGEN is accurate. Since CMFGEN provides a good fit to the photospheric spectrum, there is no a priori reason that CMFGEN should get the ionization structure dramatically wrong. Studies of H\,{\sc ii} regions generally find that the far UV spectrum predicted by CMFGEN is globally consistent with that required to ionize the nebulae \citep{morisset04,sergio08}.
\item
CMFGEN does not compute the wind properties from first principles (see Sect.\ \ref{s_models}). The mass loss rate and velocity law are adopted as input to compute the atmosphere structure and the emerging spectra. However, the results can be used to check the consistency between the adopted mass-loss rate/velocity law and the momentum absorbed by the wind (the knowledge of the atmospheric structure allows the computation of the radiative forces).
\item
Recent developments in the understanding of the clumping properties of O stars have shown that porosity (and its effect on the velocity structure, called vorosity) could alter the shape of key diagnostic lines \citep[e.g.][]{sund10}. Neglecting this effect could thus bias our determination.

\end{itemize}

In the following we investigate in more detail the last two points. To check the hydrodynamics of our models, we have run several tests on HD~46202 which is typical of the late type O stars with only weak evidence for winds in \ion{C}{iv} 1548--1550 (and perhaps \ion{N}{v} 1240). We derive a UV mass-loss rate of 1.0 $\times 10^{-9}$\,\myr. This is a factor of 5 lower than the single-line limit (4.8 $\times 10^{-9}$\,\myr), and a factor of 190 lower than that derived using Vink's prescription for estimating the mass-loss rate (which gives \mdot=1.9 $\times 10^{-7}$\,\myr), and is primarily based on the strength of the absorption associated with the \ion{C}{iv} 1548--1550 doublet. As noted previously, the single line limit is larger than the derived UV mass loss rate. To check the hydrodynamics, we ran additional models with as many elements/lines as allowed by our computational resources (the models include H, He, C, N, O, Ne, Mg, Si, S, Ar, Ca, Fe, Ni and a total of about 10\, 000 atomic levels) to be able to calculate the radiative force as accurately as possible.

The momentum equation can be written as:

\begin{equation}
\label{eq1}
V{dV \over dr} = {-1\over \rho} {dP\over dr} - g + g_r
\end{equation}
where
\begin{equation}
g=GM/r^2
\end{equation}
and, $g_r$, the radiative acceleration, is given by
\begin{equation}
g_r = \frac{4\pi}{c} \rho \int \chi_\nu H_\nu d\nu \,.
\end{equation}

\noindent Once $g_r$ is computed, we can thus check whether Eq.\ \ref{eq1} is satisfied or not.

When clumping is important we utilize the same equation, however this is only an approximation. First we note that the presence of clumps implies a dynamic flow, but the above equation is for a steady flow. Second, it is unclear how to handle the pressure term when clumping is important -- in practice this is not crucial for O stars since we assume the clumping to be initiated close to, or above the sonic point where the gas pressure term can be neglected. The expression for $g_r$, $\rho$ and $\chi_\nu$ are evaluated in the clumps.

\begin{figure*}[!t]
     \centering
     \subfigure[HD 46150]{
          \includegraphics[width=.48\textwidth]{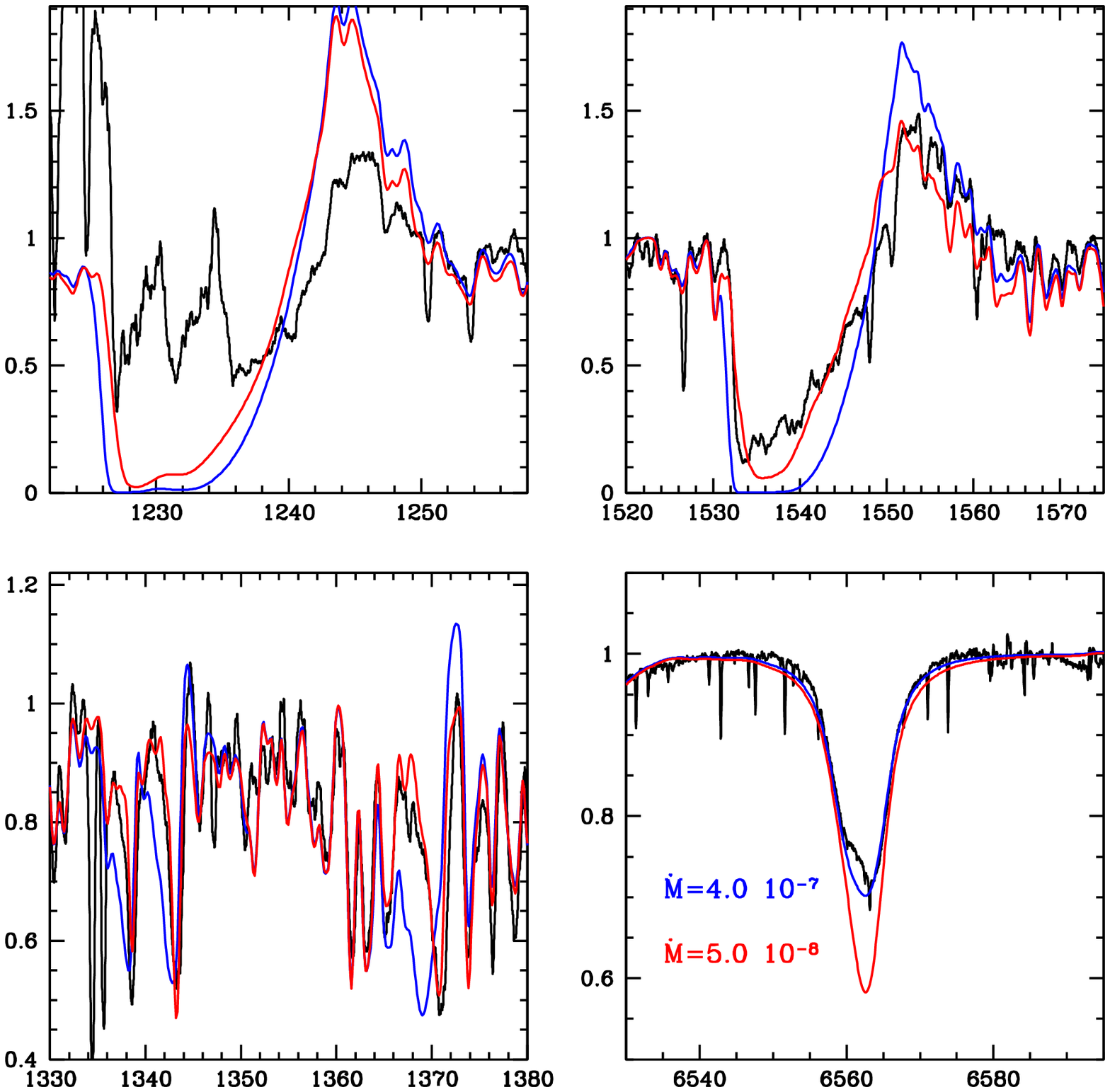}}
     \hspace{0.2cm}
     \subfigure[HD 46202]{
          \includegraphics[width=.48\textwidth]{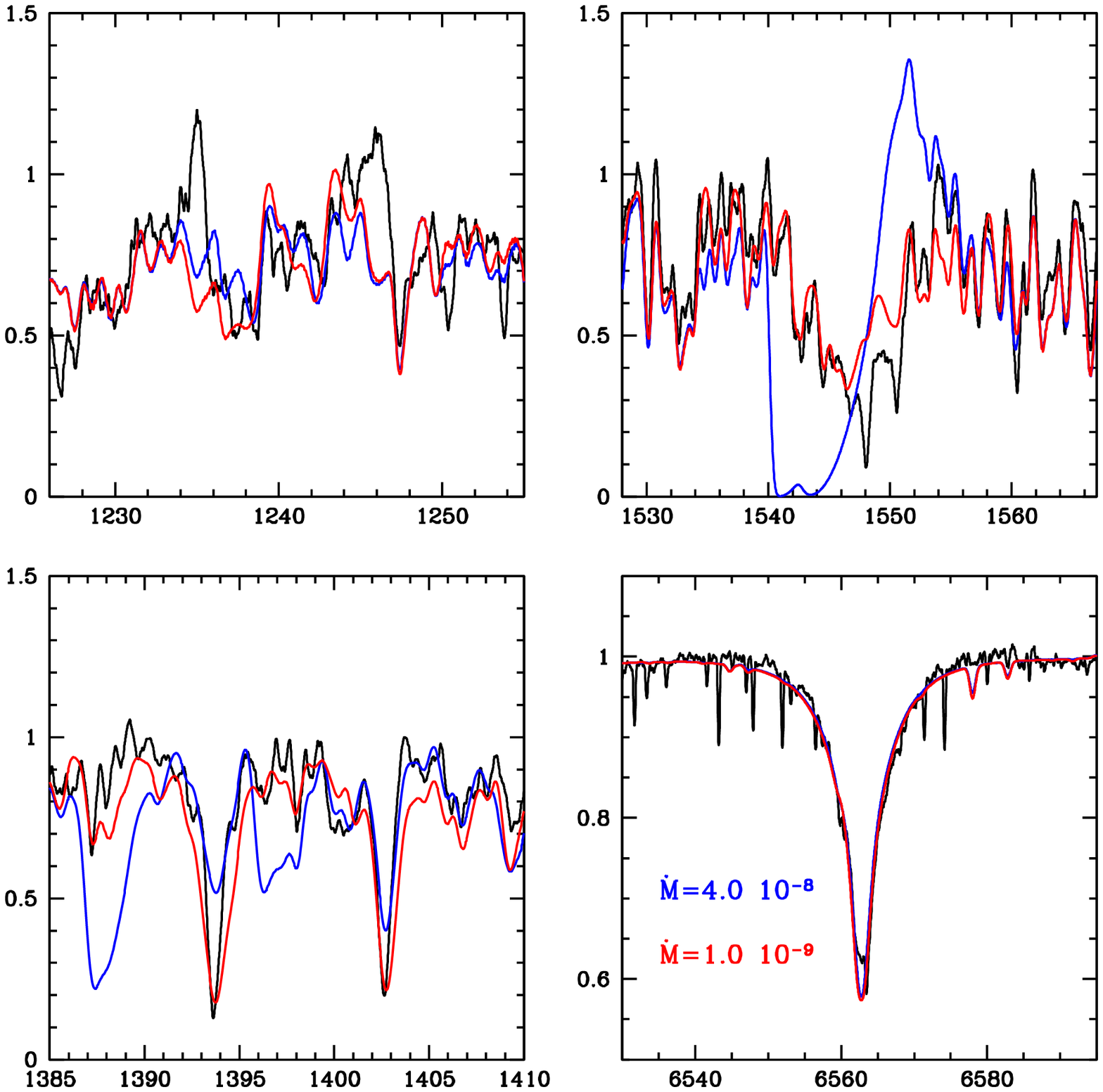}}
     \caption{Comparison between the best fit models of the UV spectrum (red lines) and models with higher mass loss rates (blue lines). The solid lines are the observed spectra. The left panel shows the comparison for HD~46150, the right panel for HD~46202. The narrow absorption features in the \ion{C}{iv} 1548--1550 profiles are of interstellar nature.}
     \label{fig_mdots}
\end{figure*}

In our best fit model to the UV lines, the radiation field exceeds that required to drive the wind everywhere -- at most locations by over an order of magnitude. This result is not unexpected, since our derived mass-loss rate is lower than the single-line limit.
As a check of the consistency of the calculations we computed a model using a mass loss rate of 4.0 $\times 10^{-8}$\,\myr -- this is a factor five below Vink's estimate, but a factor of 40 above what we derived. As shown in Fig.\ \ref{fig_mdots} (right panel), the predicted UV spectrum shows some crucial disagreements with observation.
Particularly noteworthy is the presence of a saturated C\,{\sc iv}  profile, and two high-velocity wind
features due to Si\,{\sc iv}. In the observed spectrum, Si\,{\sc iv} shows only photospheric components\footnote{If we adopted a lower \teff\ for HD~46202, the disagreement with Si\,{\sc iv} would be even more striking}. The high mass loss rate model shown has no X-rays --- when we include X-rays \ion{N}{v} 1240, unlike the observations, shows a strong P ~Cygni profile. The weakness of the N\,{\sc v} profile relative to C\,{\sc iv} is consistent with the late spectral class of HD~46202, and argues against a shift  of the wind to high ionization stages.

For a mass-loss rate of 4 $\times 10^{-8}$\,\myr, the wind momentum deposited is close to that needed
to drive the wind for velocities in excess of 700 \kms, and for a standard $\beta=1$ velocity law. Below 700\,\kms\ there is insufficient line force, sometimes by a large factor, to drive the flow. This is an important issue --- for a steady flow the momentum balance must be satisfied at all depths, not just
globally. Clearly we have an inconsistency. Either the derived UV mass-loss rate is wrong or there is an additional retarding force that needs to be included in the momentum balance equation.

For early type O dwarfs, mass loss rates are larger and closer to the theoretical predictions. However, problems still occur. We take HD~46150 as an illustrative case. For that star, we derive a mass-loss of $~5.0\times 10^{-8}$ \myr\ from UV diagnostics. 
This gives a C\,{\sc iv} profile in reasonable agreement with observation, but the N\,{\sc v} doublet
is overestimated. Surprisingly both the C\,{\sc iv} and N\,{\sc v} profile do not reach zero intensity (C\,{\sc iv} has a residual intensity of ~15\%  while the residual intensity for N\,{\sc v} is ~35\%).

On the other hand, the mass-loss derived from H$\alpha$ (assuming $f$=0.1, V$_{cl}$=100 km/s) is nearly an order of magnitude larger -- $~4.0\times 10^{-7}$ \myr.  This mass--loss rate also provides a better fit to the cores of H$\beta$ and H$\delta$. 
The optically derived mass-loss is inconsistent with the \ion{O}{iv} 1339,1343  and \ion{O}{v} 1371 lines -- the theoretical spectra show noticeable perturbations by the wind whereas the observed profiles are photospheric only (see Fig.\ \ref{fig_mdots}).  The fit to \ion{C}{iv} is somewhat worse although definitive statements are impossible since \ion{C}{iv}  shows a strong P~Cygni profile that is insensitive to \mdot.

Clumping, and more precisely porosity, might be a way to partly solve the issues mentioned above. Porosity comes in two flavors --- spatial and velocity. Spatial porosity has been invoked to help explain the observed X-ray profiles, which tend to show much more symmetric profiles than expected on the basis of H$\alpha$ derived mass-loss rates \citep{oskinova07}. An alternative explanation is that H$\alpha$ mass-loss rates are too high --- by a factor of a few \citep{cohen10}. The need for spatial porosity, and its importance, is controversial. \citet{cohen10} argue that the the X-ray profiles in $\zeta$~Pup are much better explained by a simple reduction in mass-loss. In addition, \citet{oc06} have argued that the wind clumping structure is inconsistent with that required to give rise to a significant porosity effect.
 
Vorosity (i.e. porosity in velocity space) can arise in a clumped medium because the clump velocities are discrete -- along a given sight line not all velocities may be present. Since continuum transfer is unaffected by the presence of velocity fields, vorosity, unlike spatial porosity, only affects line profiles.\footnote{For the X-ray lines vorosity is not important -- it is the continuum opacity which primarily affects the shape of these lines.} As shown by \citet{pm10}, vorosity could explain that the ratio of optical depth in the two components of \ion{Si}{iv} 1393,1403 in B supergiants is not equal to the ratio of oscillator strengths, as expected in the case of homogeneous flows.

\begin{figure*}[!t]
\centering
\includegraphics[width=19cm]{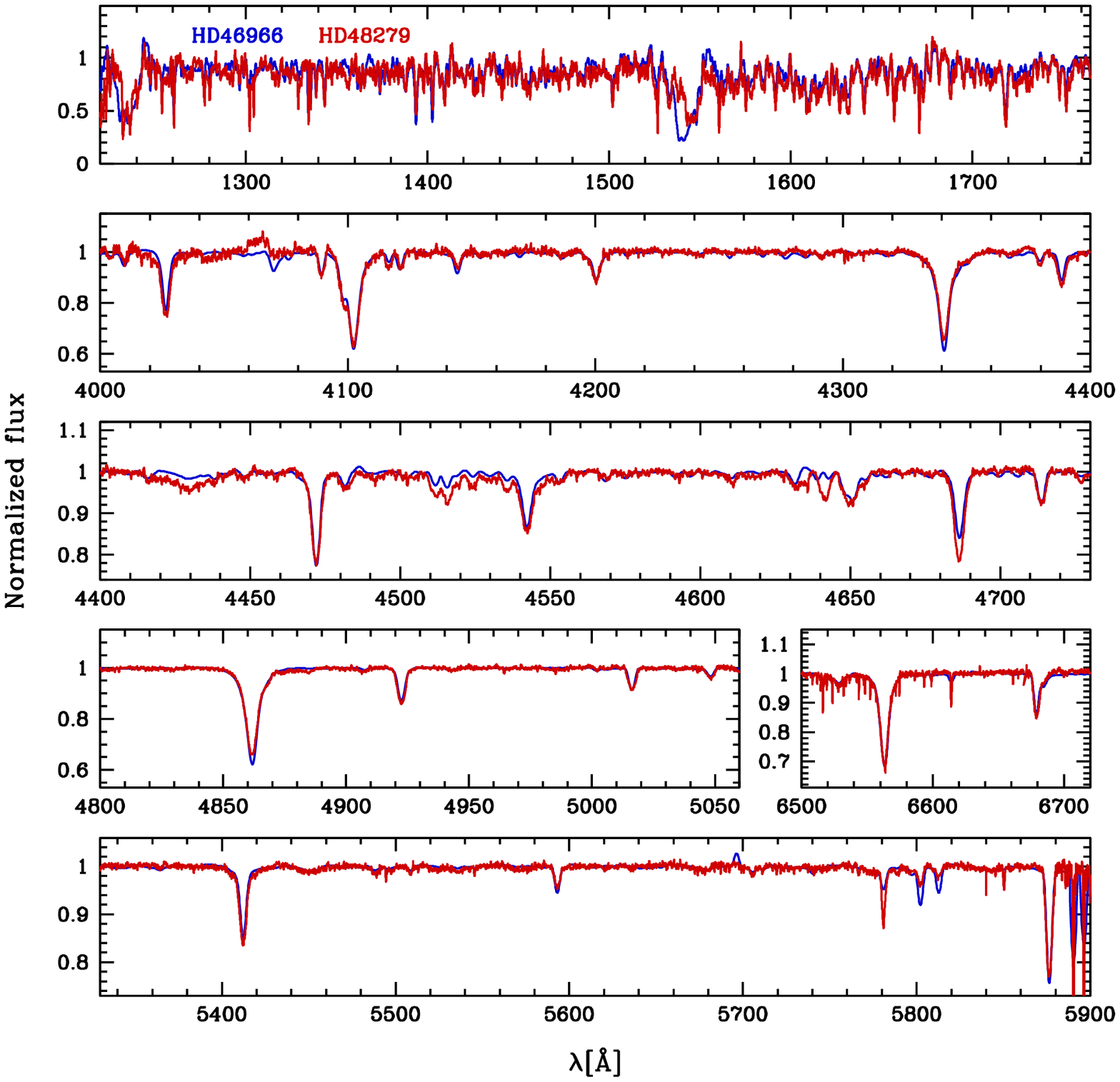}
\caption{Comparison between the observed spectra of HD46966 (blue) and HD48279 (red). The spectra of HD46966 have been convolved to the same rotational velocity as HD48279. The similar effective temperature and gravity in both stars is obvious from the \ion{He}{i}, \ion{He}{ii} and Balmer lines. On the contrary, HD46966 presents a stronger \ion{C}{iv} 1548--1550 line (but a similar H${\alpha}$ profile) and a lower N/C ratio as clearly seen from the \ion{C}{iii} 4070, \ion{C}{iii} 4634--4640, \ion{C}{iii} 5696, \ion{C}{iv} 5801--5812 and \ion{N}{iii} 4500--4520 lines.} \label{comp_midO}
\end{figure*}

Recent simulations by \citet{sund10,sund11} have highlighted the importance of vorosity (among others) on the formation of strong UV lines. They show that accounting for vorosity, UV line profiles can be strongly reduced for a given mass loss rate. At the same time, H${\alpha}$ is less affected. This tends to bring the optical and UV mass loss rates into better agreement when fitting oberved spectra. But these results also indicate that if vorosity is important, it implies that the winds are clumped. Consequently the H$\alpha$ mass loss rates derived using homogeneous models (such as most values in Table \ref{tab_lc2}) are upper limits. These computations are still exploratory (there is no detailed radiative transfer nor feedback on the atmosphere structure) but they appear to be a way to solve the issues presented above.

Aside from this, the effect of porosity and vorosity on theoretical mass-loss rates is unclear. If it begins beyond the critical point we would expect a reduction in the line force, and hence a reduction in the terminal velocity of the wind, with no change in the mass-loss rate. The reduction in line force occurs because thick lines would absorb less continuum flux, due to the presence of absorbing material only at specific velocities and not all over the entire 0--\vinf\ velocity range. The radiation force from thin lines would not alter since only the column density, not the velocity field, is important. Of course the definition of which lines are thin and which are optically thick may be affected by the wind structure. If vorosity is important below the critical point we might expect a reduction in the mass-loss rate.

Clearly, the structure, properties and dynamics of O star winds is complex and requires improvements in both spectroscopic analysis and theoretical predictions. But vorosity appears as a valid way of solving (at least partly) some of the issues raised above.

Whatever the exact nature of these winds, some qualitative features can be obtained directly from the observations. This is what we show in Fig.\ \ref{comp_midO}. Most of the results presented above rely directly or indirectly on atmosphere models. In Fig.\ \ref{comp_midO} we show a direct comparison between the observed spectra of two stars: HD46966 and HD48279. According to Table \ref{tab_param}, they have very similar \teff\ and gravities. This is mirrored in Fig.\ \ref{comp_midO} by the perfect match of the \ion{He}{i} and \ion{He}{ii} features and the wings of Balmer lines. However, \ion{N}{v} 1240 and \ion{C}{iv} 1548--1550 show a smaller blueward extent in HD48279 indicating a lower terminal velocity. In addition,  \ion{C}{iv} 1548--1550 is clearly weaker in HD48279 while the \ion{N}{v} 1240 profile are of similar strength (slightly weaker in HD48279 though). The H${\alpha}$ profiles are barely differenciated. From the \ion{C}{iii} optical features, we tentatively estimate the carbon abundance (C/H) of HD48279 (respectively HD~46966) to be $1.0 \times 10^{-4}$ (resp. $2.0 \times 10^{-4}$). Consequently, HD~48279 exhibits a larger N/C ratio compared to HD46966. If the mass loss rate were the same in both stars, one would expect a weaker \ion{C}{iv} 1548--1550 line but also a much stronger \ion{N}{v} 1240 line (the profile is not saturated) since the nitrogen overabundance in HD48279 is much larger than its carbon depletion (see Table \ref{tab_param} and above). This is not what we see. We thus conclude that the wind of HD48279 is weaker than that of HD46966. The value of the mass loss rate is low enough that the H${\alpha}$ profiles are photospheric in both stars, and thus similar since \teff\ and \logg\ are the same. HD46966 has a luminosity $L=10^{5.20}L_{\sun}$ while for HD48279, $L=10^{4.95}L_{\sun}$. It is thus natural to have a weaker wind in HD48279. This is what both the UV diagnostics and H${\alpha}$ indicate. We note however the difference in mass loss rate between the two stars is 0.8 dex if the former are used, while it is only 0.4 dex in the case of H${\alpha}$. This raises once again the question of the origin of the differences between the UV and optical wind sensitive lines. Whatever the reason, we see that two stars with very similar effective temperature and gravity can have different wind properties with only a difference in luminosity by less than a factor of two.

\subsection{Chemical evolution and rotation}
\label{s_hunt}

\begin{figure}[!t]
\centering
\includegraphics[width=9cm]{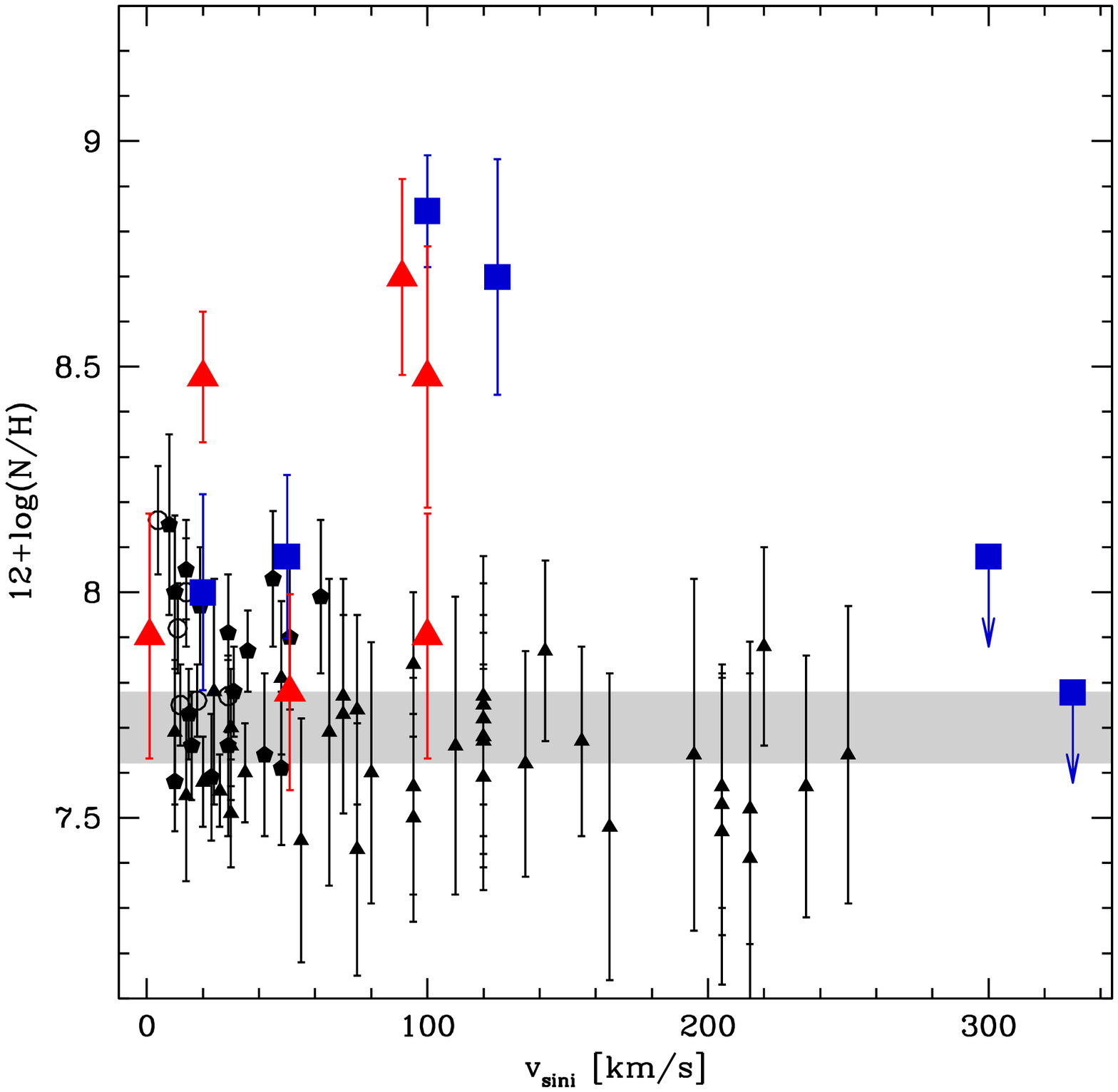}
\caption{Nitrogen surface abundance (in units of 12+log(N/H)) as a function of projected rotational velocity. Single O stars are the blue squares and binaries the red triangles. The small black triangles (pentagons, open circles) are the Galactic B stars of \citet{hunter09} \citep{morel08,przy10} with \logg\ larger than 3.5 (i.e. comparable to our sample). The grey area corresponds to the range of solar values from studies of the Sun, B stars and HII regions \citep[see Table 3 of][]{hunter09}.} \label{hunt}
\end{figure}

\begin{figure}[!t]
\centering
\includegraphics[width=9cm]{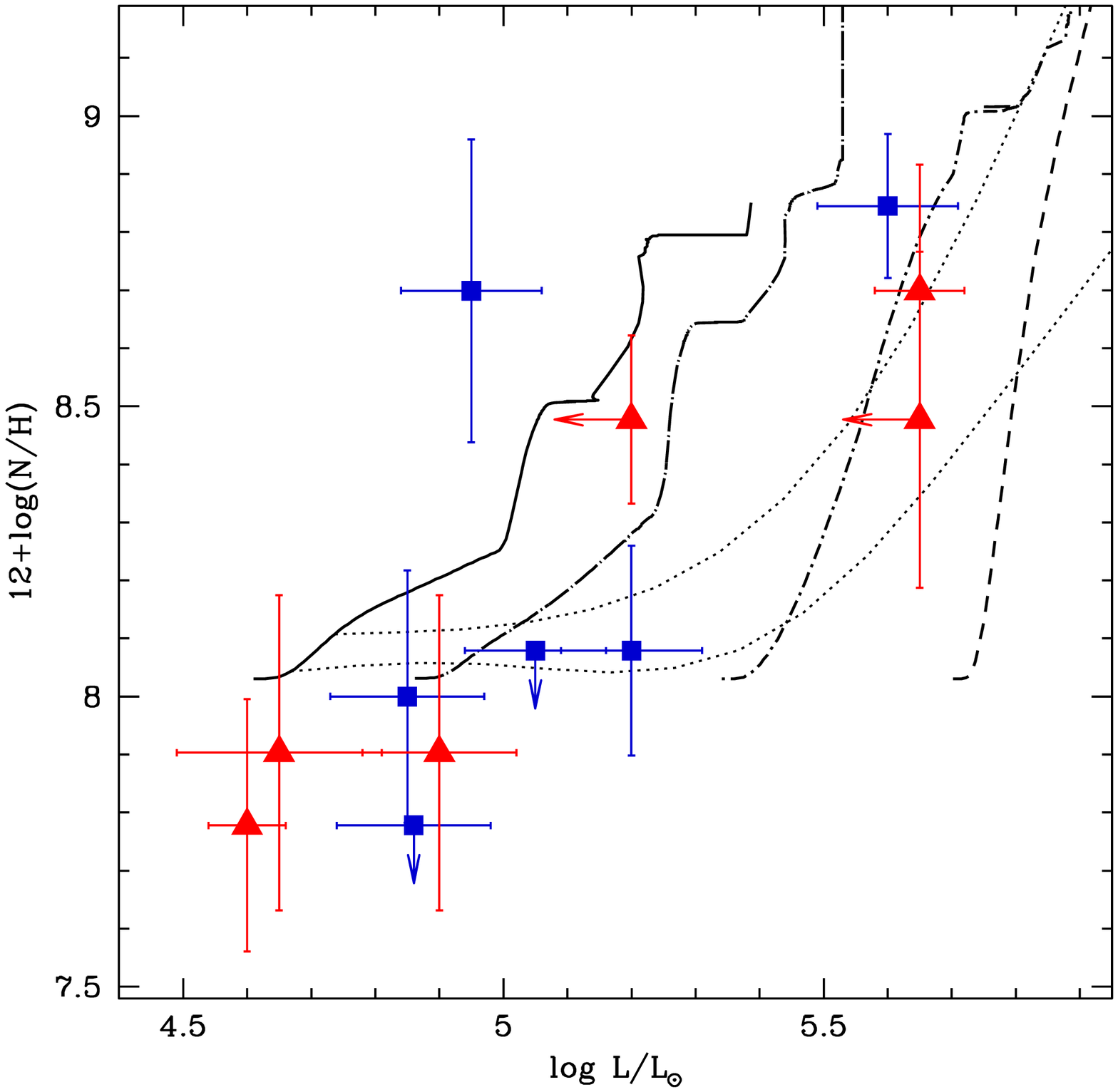}
\caption{Nitrogen surface abundance (in units of 12+log(N/H)) as a function of luminosity. Single O stars are the blue squares and binaries the red triangles. Evolutionary tracks -- shown by the solid, dot-long dashed, dot-dashed, dashed lines for $M$=15, 20, 25 and 40 \msun\ -- are from \citet{mm03}. The dotted lines are isochrones for 2 and 4 Myr.} \label{N_L}
\end{figure}

Rotation is known to affect the angular momentum and chemical elements transport through mixing processes \citep[meridional circulation and shear turbulence, see][]{mm00}. The first analysis of surface CNO abundances of O stars and B supergiants tends to confirm qualitatively the theoretical predictions: more evolved stars have a higher N content as a result of CNO processing and mixing \citep{hil03,heap06,paul06bsg}. However, the role of rotation in explaining the chemical patterns of B stars in the Magellanic Clouds has been questioned by the results of the first ESO/VLT Large Programme dedicated to massive stars \citep{evans05,evans06}. \citet{hunter08} showed that 60\% of their sample of B main--sequence stars in the Large Magellanic Cloud displayed nitrogen surface abundances expected by evolutionary models including rotational mixing, but the remaining 40\% escaped such an agreement. These outliers are divided in two groups with approximately the same number of members: one with highly enriched but slowly rotating stars; the other with barely enriched fast rotators. The former group was also identified in the SMC \citep{hunter09}. In the Galaxy, the results of \citet{hunter09} point to a lack of surface nitrogen enrichment in all main--sequence objects, probably due to the relatively unevolved status of most stars. In contrast, \citet{morel06,morel08} reported the detection of N-rich non-supergiant Galactic B stars with low rotation rates, lending support to the idea that rotational mixing cannot explain the surface abundances of a significant number of massive stars also in the Galaxy.

However, \citet{maeder09} argued that surface chemical enrichment is a multivariate function and does not depend only on projected rotational velocity. The higher the mass of a star, the larger the nitrogen enrichment. Similarly, lower metallicity stars have larger surface enrichment at a given evolutionary state. In addition, more evolved stars show larger N/H ratios. Hence, a combination of ages, masses, metallicities and rotational velocities can lead to various surface enrichments. It is important to disentangle the different effects to draw reliable conclusions. According to \citet{maeder09}, if this is done, rotational mixing explains most of the observed abundance patterns. 20\% of the sample of \citet{hunter08} might escape this scenario due to binary evolution. \citet{brott11b} built population synthesis models to take all these effects into account. They investigated the properties of a population of 7--60 \msun\ stars (with a Salpeter IMF) formed continuously for the last 30 Myr and with rotational velocities drawn from the distribution of \citet{hunter08}. They confirm that about 35\% of the stars observed by \citet{hunter08} cannot be explained by single star evolutionary models with rotation. But the remaing -- and majority -- of the population is well accounted for by rotating evolutionary models. \citet{przy10} also confirm the role of rotational mixing by showing a tight correlation between N/C and N/O in main sequence and supergiants stars. However, the theoretical enrichments are too large compared to observations. 

Most of the results summarized above have been obtained from the analysis of surface abundances of B main--sequence stars. In this paper, we present results for main--sequence O stars. Our sample is smaller than those of \citet{morel08} or \citet{hunter08} but it allows to identify some trends, and extend their work to the most massive stars. Fig.\ \ref{hunt} shows the nitrogen surface abundance as a function of rotational velocity. The results of \citet{morel08}, \citet{hunter08} and \citet{przy10} have been added for comparison (small symbols). The following comments can be made:

\begin{itemize}

\item[$\bullet$] up to about 120 \kms\ a possible trend of larger N/H with larger \vsini\ is emerging among O stars, although a significant scatter exists and more stars are definitely needed to draw firm conclusions. We note that our sample is relatively homogeneous in age and metallicity. On the other hand it covers a relatively wide mass range (15--60 \msun). Both binaries and single stars seem to follow this trend.

\item[$\bullet$] the components of the binaries studied in this paper have surface N/H values consistent with those of single stars. This conclusion applies only to the binaries studied here and should not be generalized.

\item[$\bullet$] the two fastest rotators show very little enrichment if any. 

\item[$\bullet$] on the main--sequence, in the Galaxy, O stars show on average larger values of N/H than B stars.

\end{itemize}

\noindent These results confirm some of the expectations of stellar evolution with rotation. First, more massive stars display larger surface nitrogen enrichment. This is even more clearly illustrated on Fig.\ \ref{N_L} where the ratio N/H is shown as a function of luminosity. A clear trend of higher N content for more luminous stars is observed. At a given age, this is what evolutionary models predict. The dotted lines show isochrones for 2 and 4 Myr, typical of the age of our sample stars. They indicate faster evolution and consequently faster chemical enrichment for more luminous and thus more massive stars. Interestingly, the two fast rotators do not appear as outliers in this plot, but fall perfectly in the sequence followed by most of the stars. This means that given their mass, these objects are too young to have experienced a strong enrichment in spite of their large rotational velocity. This confirms the conclusion of \citet{maeder09} that it is important to disentangle the various effects at work in the surface enrichment of massive stars to pin down the true role of rotation. 

Our sample includes binaries. In Fig.\ \ref{hunt} and \ref{N_L} they do not show any peculiar property in terms of chemical enrichment compared to single O stars. This does not mean that binarity has no effect on the history of surface enrichment, but that it depends on the binary properties. In particular, not all binaries have experienced mass transfer which would affect both the rotational speed and surface composition \citep{linder08,demink09}. With the exception of HD~48099, all the binaries of our sample have orbital periods larger than 3 days. According to \citet{demink11}, it takes more than 8 Myr for Roche--lobe overflow to occur in such systems. This result is valid for binaries of about 20 \msun\ and obviously depends on masses and mass ratios. The binaries of our sample are thus most likely too young and/or too separated to have experienced significant mass transfer. This conclusion only applies to the systems included in our sample and should not be generalized to all binary stars. Once mass transfer has occured, it is likely that additional mixing processes take place and affect the surface composition. Further analysis of such interacting systems should shed more light on the exact role of binarity on the chemical appearance of O stars.

In Fig.\ \ref{N_L}, the only clear outlier is star HD48279 with a very strong enrichment in spite of a moderate luminosity. To a lesser extent, HD46573 can also be considered rather enriched compared to the expectations of single star evolutionary models due to the upper limit on its luminosity. For HD48279 it does not seem possible that rotation only can explain the large N content. According to Fig.\ 1 of \citet{maeder09}, an increase of the rotational velocity from 150 \kms\ to 300 \kms\ leads to an extra enrichment of the order 0.5 dex. HD48279 is 0.7 dex more N rich than the other stars of the same luminosity in Fig.\ \ref{N_L}. Binarity is excluded for that object \citep{mahy09}. The only possible explanation to the strong N enrichment is that HD48279 was member of a binary system in the past and experienced mass transfer and/or tidal interaction. Either the binary has been disrupted (by supernova kick for instance) or the companion is a low--mass compact object rendering any radial velocity variation undetectable \citep[see][]{demink11}. In the case of HD46573, a large initial velocity or binarity can explain the possible slightly larger N/H ratio compared to stars of the same luminosity range \citep[HD46573 has been classified as SB1 by][]{mahy09}. Finally, HD46150 also only has an upper limit on its luminosity. But even accounting for an extreme reduction by 0.3 dex, the N/H content would still be consistent with single star evolutionary tracks.

\subsection{Stellar masses}
\label{s_mass}

\begin{figure}[!t]
\centering
\includegraphics[width=9cm]{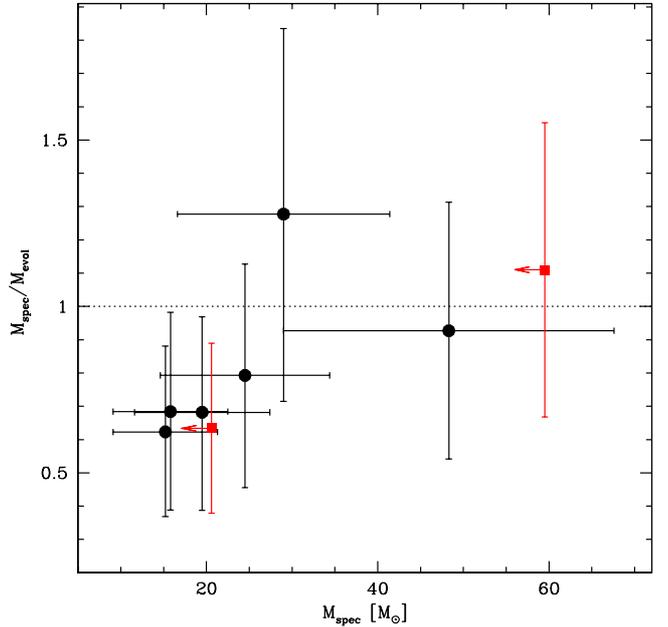}
\caption{Ratio of spectroscopic to evolutionary masses versus spectroscopic masses for the sample stars. The dotted line is the one to one relation. Black circles are single stars, red squares are members of binary systems. The insert shows the position of the primary component of HD48099.} \label{mass}
\end{figure}

The determination of stellar masses is a long--standing problem. The so-called ``mass discrepancy'' \citep{gro89,her92} refers to the systematic overestimate of evolutionary masses compared to spectroscopic masses. The former are masses derived from the HR diagram and evolutionary tracks. The latter are the masses resulting from the knowledge of the gravity and radius obtained from spectroscopic analysis with atmosphere models. The revision of the effective temperature scale of O stars \citep{msh02} has lead to improvements \citep[see][]{gies03}. However, differences of up to a factor of three still exist in some cases. \citet{massey05} and \citet{mokiem07} reported a mass discrepancy for LMC stars. The latter indicate that part of the discrepancy could be caused by fast rotation in He-rich stars. Fast rotating stars tend to evolve towards larger luminosities for the same initial mass as non rotating stars. Consequently, using non rotating tracks to analyze them leads to overestimates of their evolutionary masses. However, even non enriched objects (hence not likely to rotate rapidly) show the mass discrepancy. 

In Fig.\ \ref{mass} we show the ratio between spectroscopic (M$_{spec}$) and evolutionary (M$_{ev}$) masses for our sample of Galactic stars. Black circles are single stars and red squares the components of binary systems. The spectroscopic masses have been computed from the effective gravity (that we determine through line fitting) corrected from the effect of centrifugal forces. However, because of the high uncertainty on the gravity of the SB2 systems, we do not include the components of HD 46149 and HD48099 in this discussion. In spite of the small number of object, the trend for a mass discrepancy at low masses (M $<$ 25 \msun) is present. At higher masses, a good agreement is observed, but the error bars are large. These results are qualitatively consistent with those of \citet{repolust04} who showed that stars with masses lower than about 50 \msun\ seemed to follow a relation parallel to the 1:1 relation in the M$_{ev}$ versus M$_{spec}$ diagram.

Recently, \citet{wv10} conducted a study of a sample of Galactic O stars and concluded that the mass discrepancy was solved. In their analysis, they mainly focused on binary components. They established the evolutionary masses of their stars by using a spectral type - mass relation. They used the \teff--scale of \citet{msh05} to determine an effective temperature and luminosity for each star and subsequently used the evolutionary tracks of \citet{mm03} to assign an evolutionary mass. They then compared these estimates to dynamical masses resulting from fits to velocity and light curves of binary systems. They found a good agreement between both types of masses (see their Fig.\ 3). However, the comparison with spectroscopic masses (not shown directly by Weidner \& Vink but available from their Table 3) still reveals that evolutionary masses are larger. Hence, it seems that at least part of the origin of the discrepancy is rooted in an underestimate of gravities.


\section{Conclusion}
\label{s_conc}

We have analyzed ten OB stars (nine O stars and one B star) in the NGC2244 cluster and MonOB2 association, including two bona-fide binaries and one candidate. Optical and UV spectroscopy have been obtained. Atmosphere models computed with the code CMFGEN have been used to derive the main stellar and wind parameters of the target stars. The main results are the following:

\begin{itemize}

\item[$\bullet$] all stars have an age between 1 and 5 Myr, with a trend of lower age for the most massive stars. Whether this is a real effect or an artifact of the method used to derive the stellar parameters is not clear at present.

\item[$\bullet$] we confirm the existence of weak winds in the latest type stars of our sample. For those stars, UV mass loss rates are lower than the single line driving limit. UV mass loss rates are systematically lower than H${\alpha}$ mass loss rates. Vorosity might be a solution to these issues, although its effects on dynamics are still unclear.

\item[$\bullet$] nitrogen surface abundances in O stars indicate a larger range of enrichment compared to B stars. A clear trend of larger enrichment in higher mass stars is observed, consistent with the prediction of evolutionary models for a population of stars of the same age. There is no clear difference between single and binary stars in our sample, most likely because the latter have not (yet) experienced mass transfer or too severe tidal effects. Fast rotating non enriched O stars are understood as too young to have had time to bring enough fresh nitrogen to their surface.

\item[$\bullet$] the so--called ``mass discrepancy'' still exists for the latest type O dwarfs. Above 25 \msun, evolutionary and spectroscopic masses are in agreement, within the uncertainties. Part of the discrepancy seems related to underestimates of the spectroscopic masses. 

\end{itemize}

\noindent We need to analyze more stars of the same age to confirm the trends observed, especially concerning the surface nitrogen enrichment. Winds of massive stars are complex, and a better understanding of clumping, both observationally and theoretically, is necessary to quantify their properties.

\begin{acknowledgements}
We thank the referee, Artemio Hererro, for useful comments.
LM and GR are supported by the FNRS (Belgium), by a PRODEX
XMM/Integral contract (Belspo) and by the Communaut\'e Fran\c{c}aise
de Belgique—Action de recherche concert\'ee (ARC)—Acad\'emie
Wallonie–Europe. LM and GR acknowledge the Minist\`ere de
l’Enseignement Sup\'erieur et de la Recherche de la Communaut\'e
Fran\c{c}aise for supporting their travels to O.H.P. DJH acknowledges
support from STScI theory grant HST-AR-11756.01.A. We also thank the
staff of Observatoire de Haute-Provence and of La Silla ESO
Observatory for their technical support.
\end{acknowledgements}

\bibliography{biblio.bib}

\begin{appendix}

\section{Best fit to the observed spectra.}
\begin{figure*}
     \centering
     \subfigure[HD 46573]{
          \includegraphics[width=15cm,height=11cm]{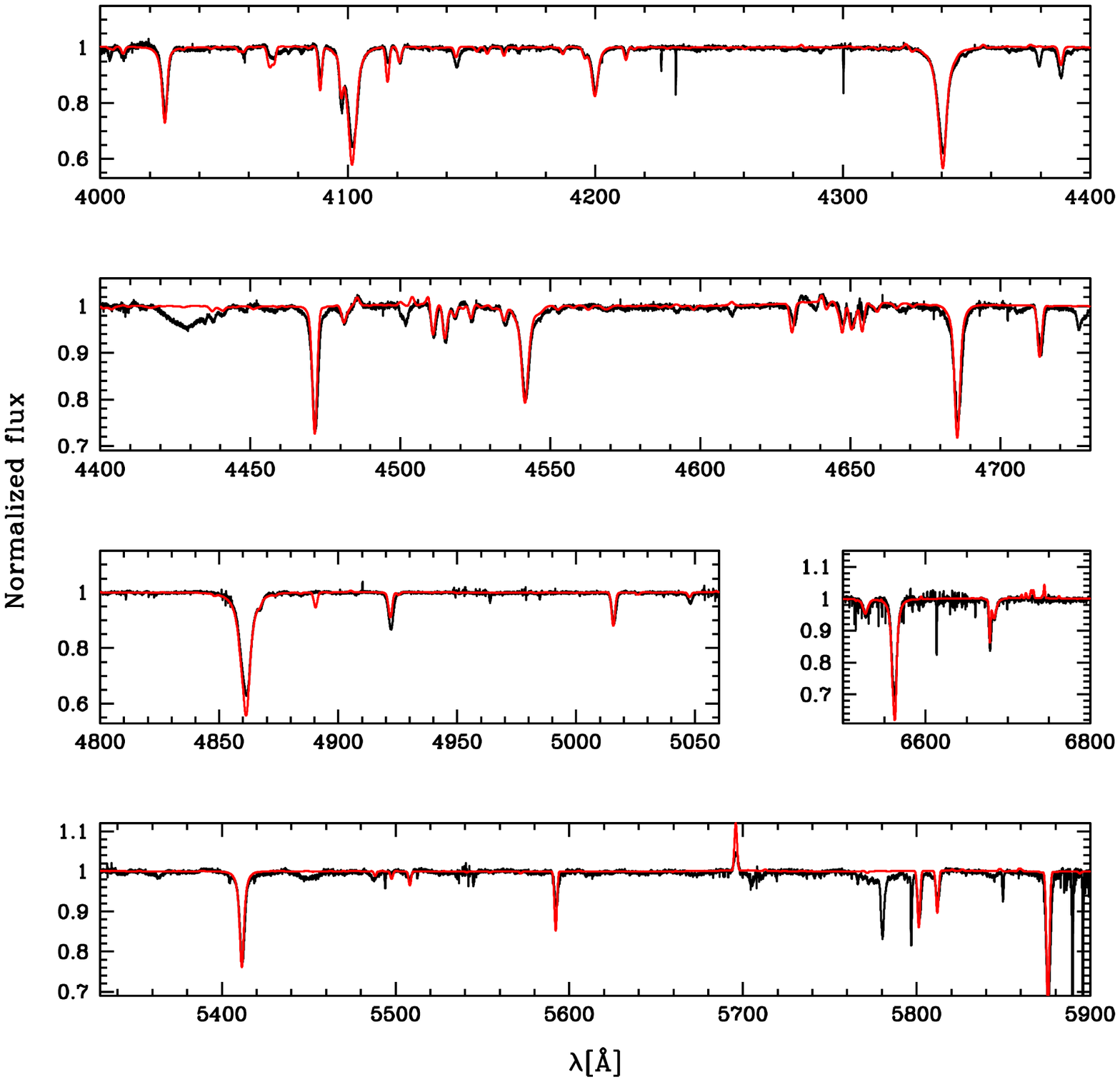}}\\
     \subfigure[HD 46485]{
          \includegraphics[width=15cm,height=11cm]{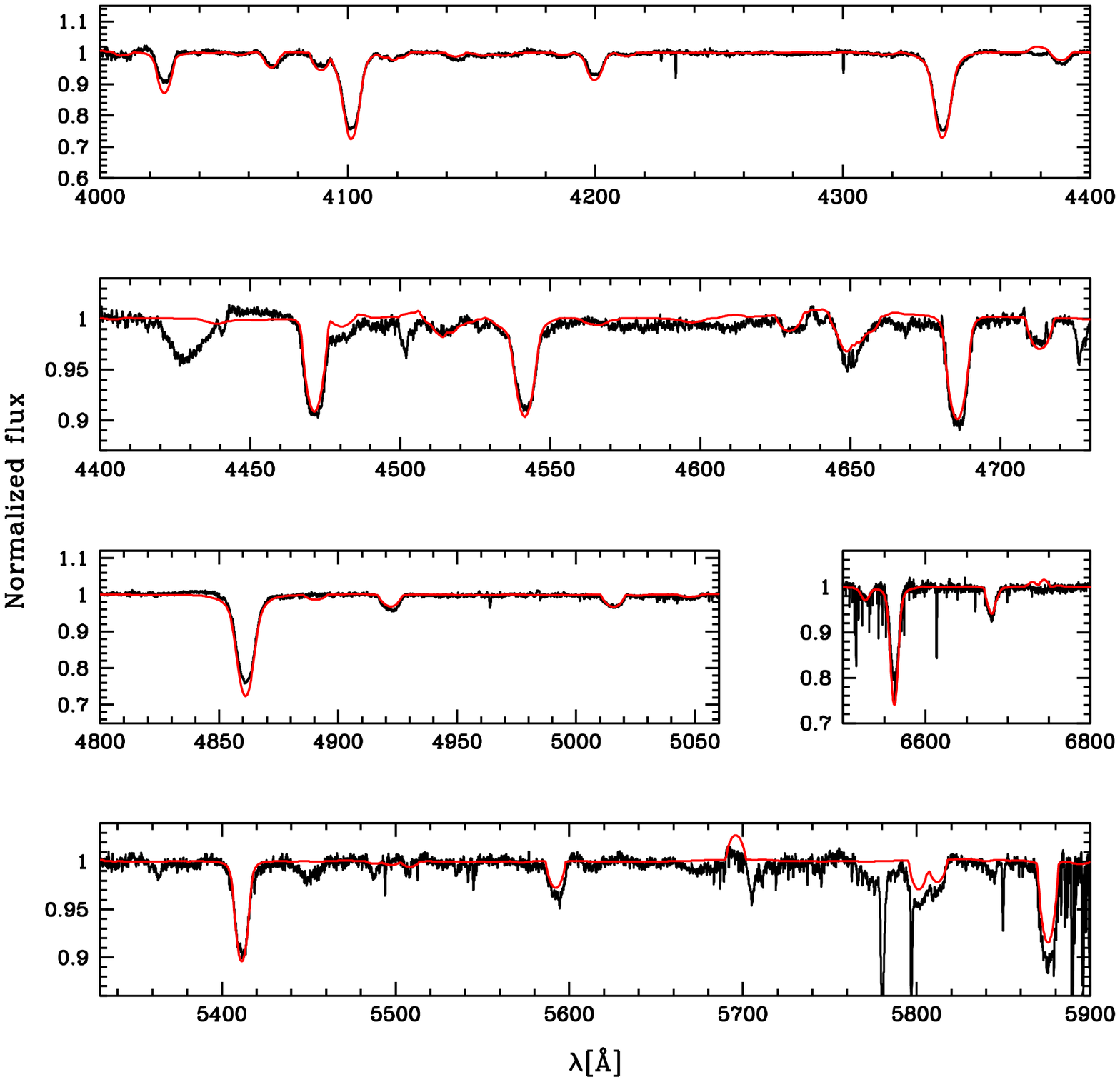}}
     \caption{Best CMFGEN fits (red) of the optical spectra (black) of stars HD46573 (top), HD46485 (bottom). The models have the mass loss rate best reproducing the UV features. The unfitted feature around 4425 \AA\ is a DIB.}
     \label{fit_A1}
\end{figure*}

\begin{figure*}
     \centering
     \subfigure[HD 46056]{
          \includegraphics[width=15cm,height=11cm]{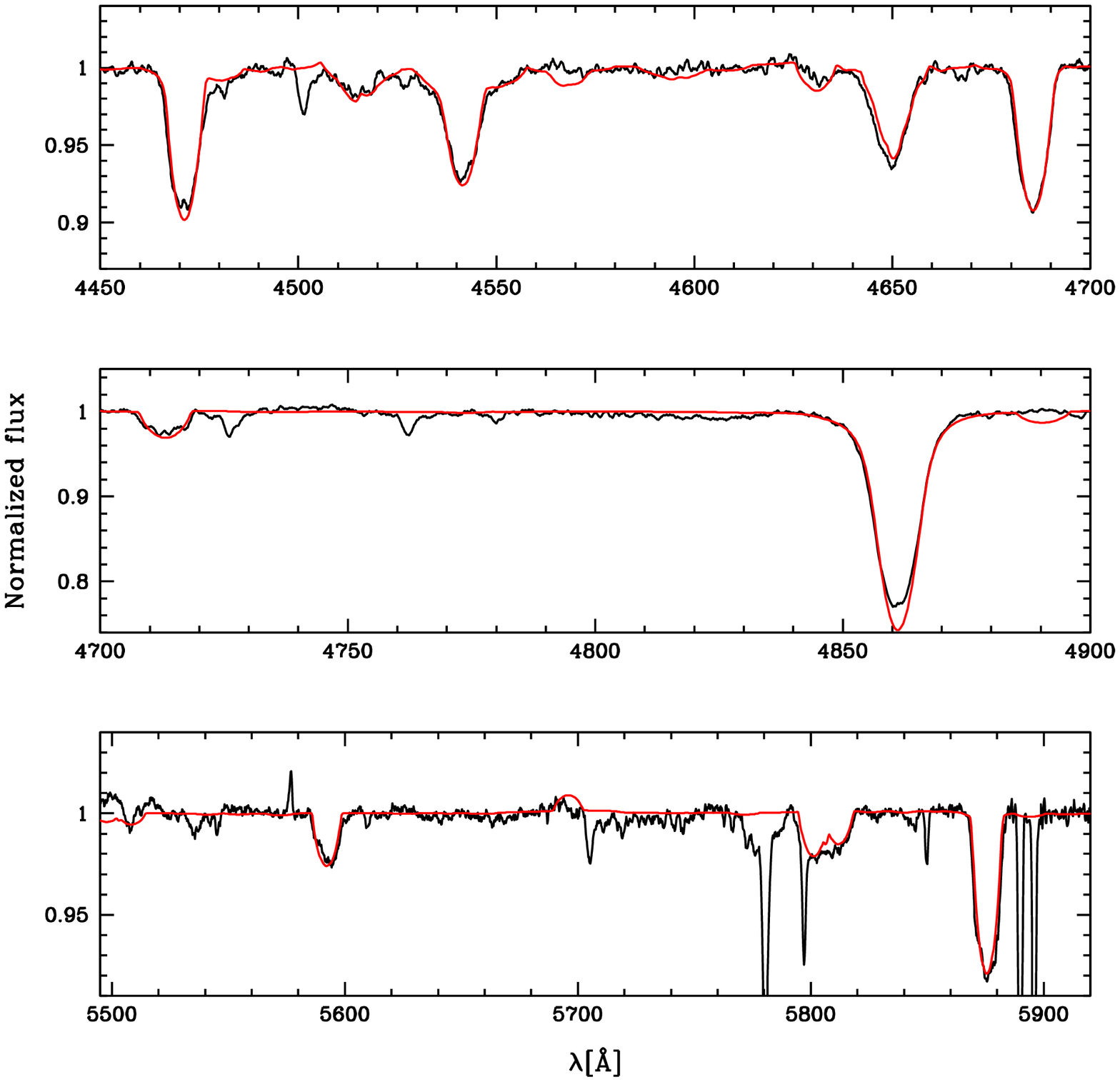}}\\
     \subfigure[HD 46966]{
          \includegraphics[width=15cm,height=11cm]{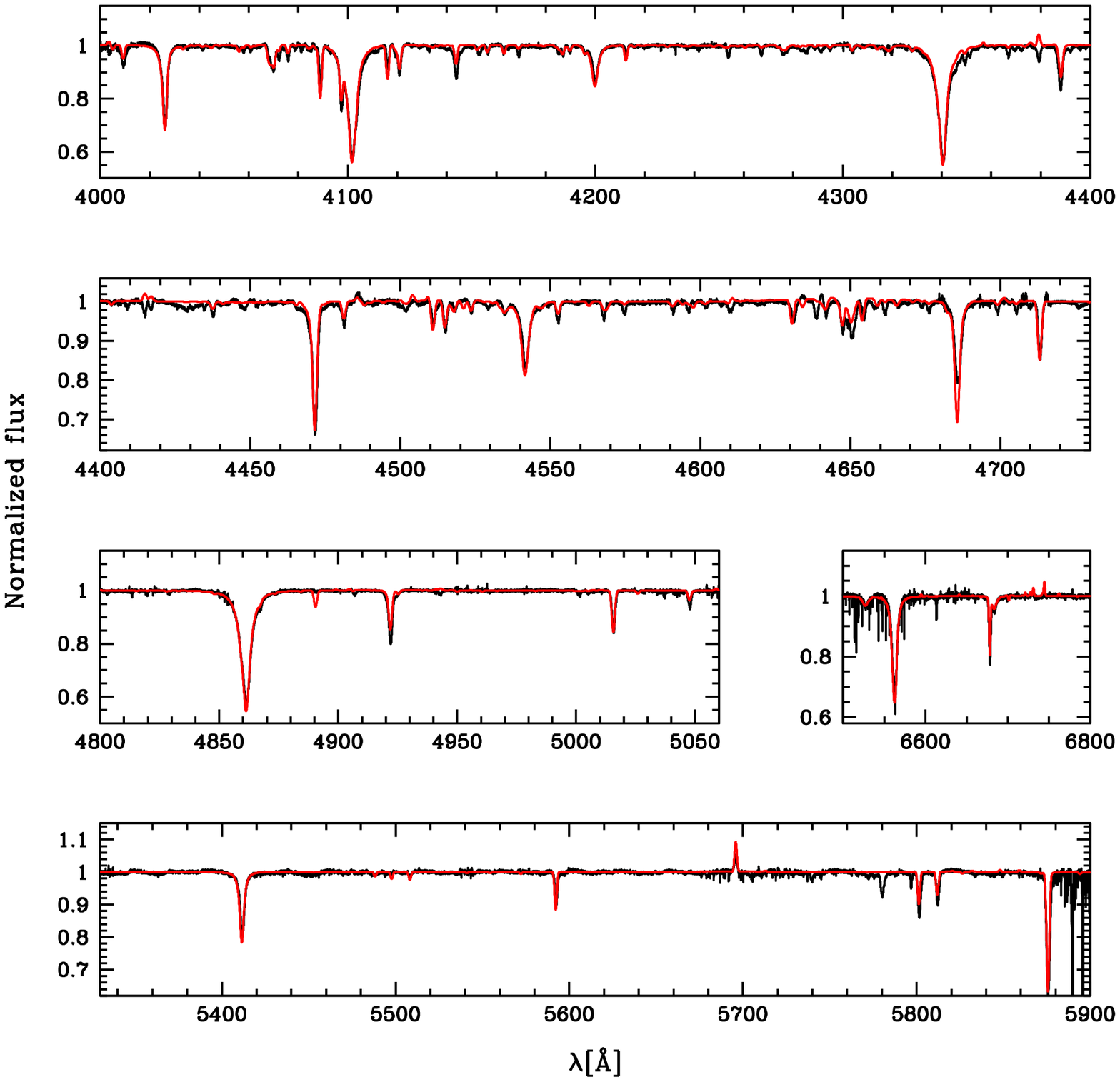}}
     \caption{Best CMFGEN fits (red) of the optical spectra (black) of stars HD46056 (top), HD46966 (bottom). The models have the mass loss rate best reproducing the UV features. The unfitted feature around 4425 \AA\ is a DIB.}
     \label{fit_A2}
\end{figure*}

\begin{figure*}
     \centering
     \subfigure[HD 46223]{
          \includegraphics[width=15cm,height=11cm]{fit_opt_hd46223.eps}}\\
     \subfigure[HD 46202]{
          \includegraphics[width=15cm,height=11cm]{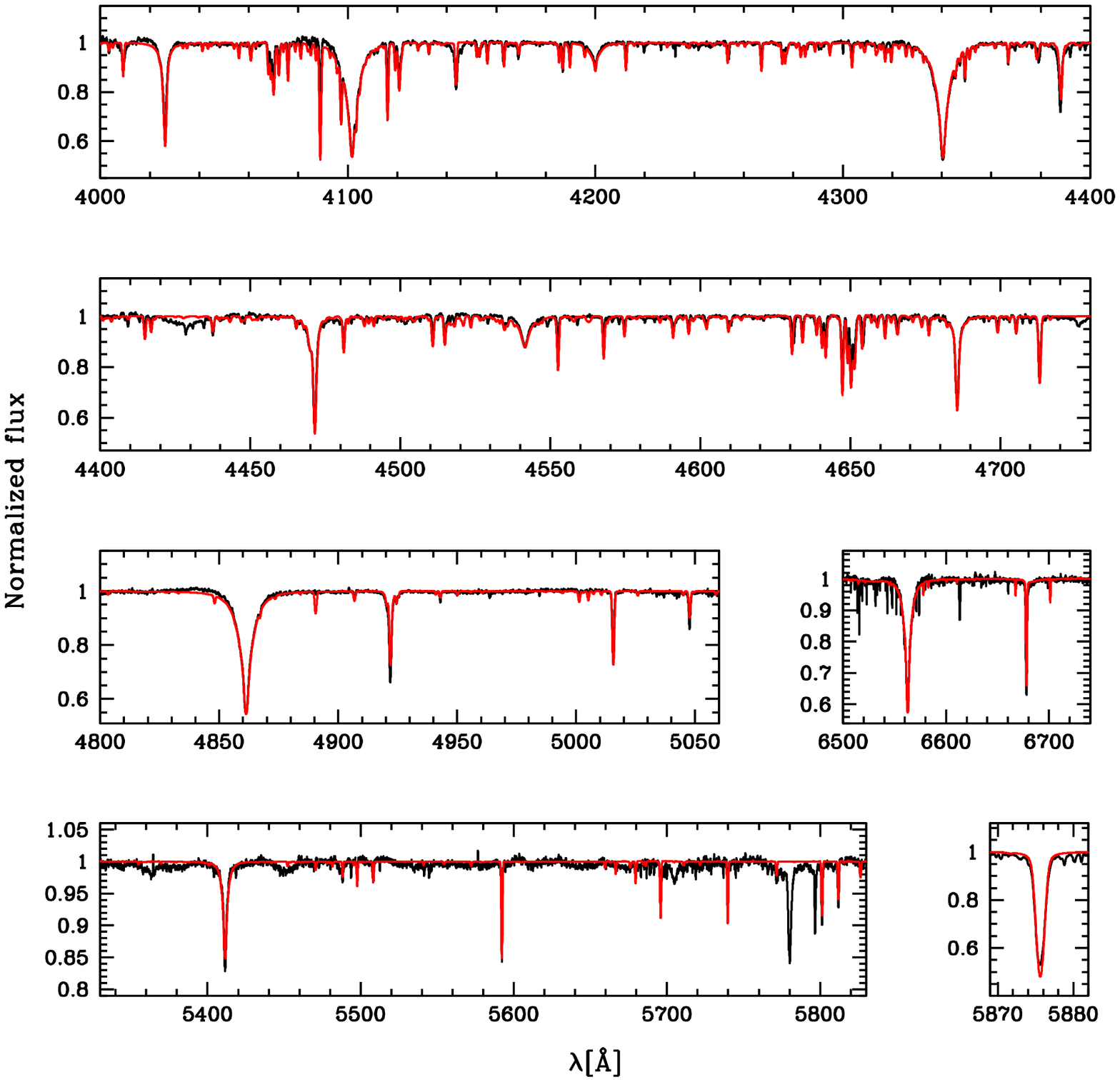}}
     \caption{Best CMFGEN fits (red) of the optical spectra (black) of stars HD46223 (top), HD46202 (bottom). The models have the mass loss rate best reproducing the UV features. The unfitted feature around 4425 \AA\ is a DIB.}
     \label{fit_A3}
\end{figure*}

\begin{figure*}
     \centering
     \subfigure[HD 46149P]{
          \includegraphics[width=15cm,height=11cm]{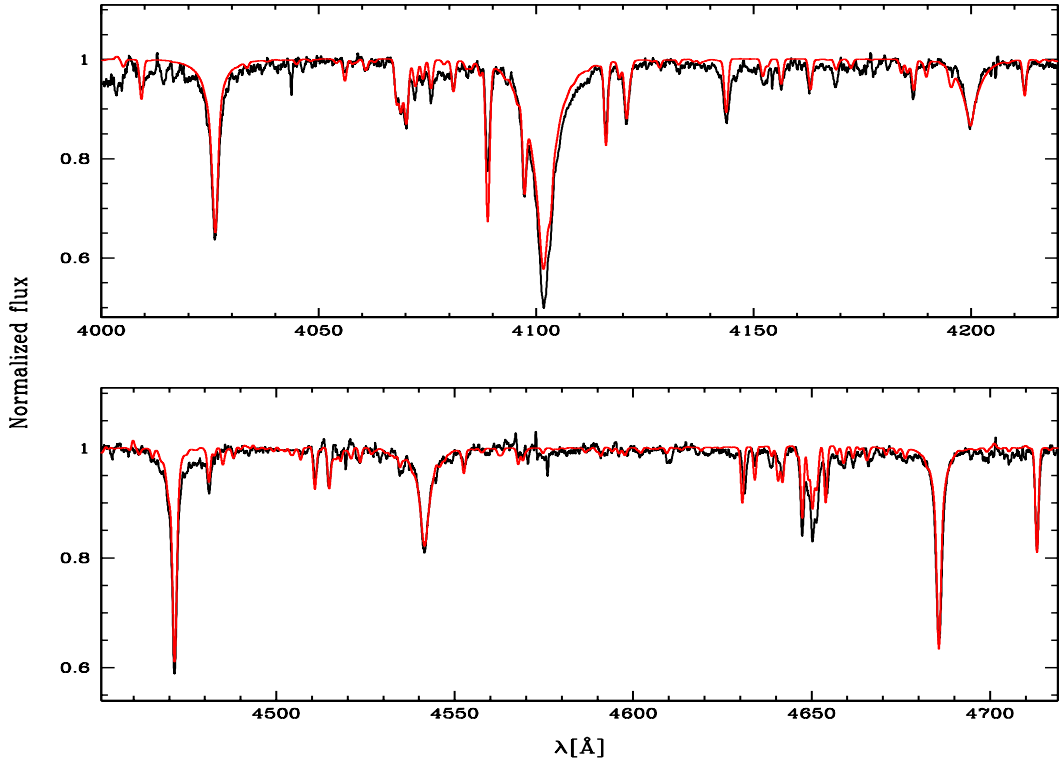}}\\
     \subfigure[HD 46149S]{
          \includegraphics[width=15cm,height=11cm]{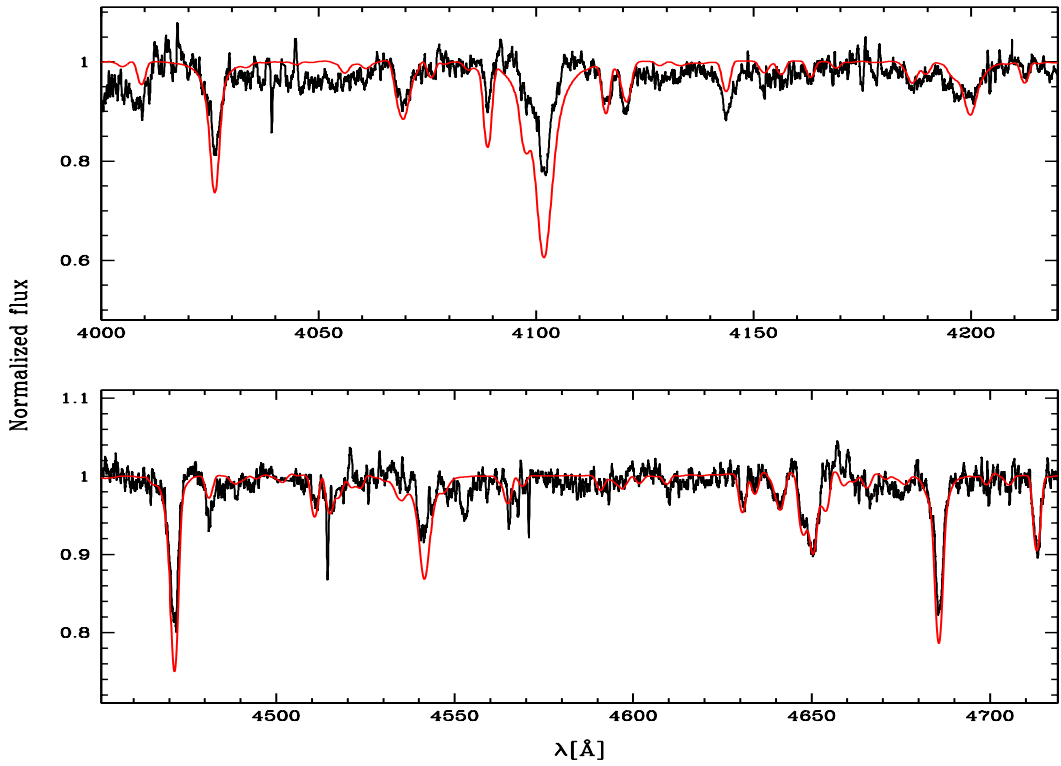}}
     \caption{Best CMFGEN fits (red) of the optical spectra (black) of the primary (top) and secondary (bottom) components of HD46149. The models have the mass loss rate best reproducing the UV features. The unfitted feature around 4425 \AA\ is a DIB.}
     \label{fit_A4}
\end{figure*}

Figures \ref{fit_A1} to \ref{fit_A4} show the observed spectra in black together with the best fit CMFGEN models in red.

\end{appendix}

\end{document}